\documentclass[a4paper,useAMS,usenatbib]{mnras}

\usepackage{graphics,epsfig,psfig}
\usepackage{todonotes}
\usepackage[normalem]{ulem}
\usepackage{xcolor}
\usepackage{float}
\usepackage{dcolumn}
\usepackage{kantlipsum,widetext}
\usepackage[style=base, singlelinecheck=off, font=small, labelfont=bf, 
justification=raggedright]{caption}
\usepackage{subfigure}
\usepackage[]{inputenc,amssymb}
\usepackage{natbib}
\usepackage{url}
\usepackage{orcidlink}
\usepackage{widetext}
\usepackage{amsmath}
\usepackage{cancel}

  \makeatletter
\def\@fnsymbol#1{\ensuremath{\ifcase#1\or \dagger\or \ddagger\or
   \mathsection\or \mathparagraph\or \|\or **\or \dagger\dagger
   \or \ddagger\ddagger \else\@ctrerr\fi}}
    \makeatother
\def \be{\begin{equation}}
\def \ee{\end{equation}}
\def \bea{\begin{eqnarray}}
\def \eea{\end{eqnarray}}

\def\aap{A\&A}
\definecolor{webgreen}{rgb}{0,.5,0}
\definecolor{webbrown}{rgb}{.6,0,0}

\def\aap{A\&A}

\def\apjs{ApJS}
\def\apjl{ApJL}

\usepackage{lineno}

\title[Spectral and Spatial Anisotropy in SGWB]{Imprints of Supermassive Black Hole Evolution on the Spectral and Spatial Anisotropy of Nano-Hertz Stochastic Gravitational-Wave Background}

\author[Sah et al.]{\fontsize{10}{12}\selectfont Mohit Raj Sah$^{1}$\thanks{mohit.sah@tifr.res.in}\orcidlink{0009-0005-9881-1788},
Suvodip Mukherjee$^{1}$\thanks{suvodip.mukherjee@tifr.res.in}\orcidlink{0000-0002-3373-5236},
Vida Saeedzadeh$^{2}$\orcidlink{0009-0000-7559-7962
},Arif Babul$^{2,3}$\orcidlink{0000-0003-1746-9529
}, Michael Tremmel$^{4}$\orcidlink{0000-0002-4353-0306},Thomas R. Quinn$^{5}$\orcidlink{0000-0001-5510-2803}\\$^{1}$Department of Astronomy and Astrophysics, Tata Institute of Fundamental Research, Mumbai 400005, India\\$^{2}$Department of Physics and Astronomy, University of Victoria, 3800 Finnerty Road, Victoria, BC, V8P 1A1, Canada\\$^{3}$Infosys Visiting Chair Professor, Indian Institute of Science, Bangalore 560012, India\\$^{4}$School of Physics, University College Cork, College Road, Cork T12 K8AF, Ireland\\$^{5}$Astronomy Department, University of Washington, Box 351580, Seattle, WA, 98195-1580, USA}

\pubyear{2024}

\begin{document}

\label{firstpage}
\pagerange{\pageref{firstpage}--\pageref{lastpage}}
\maketitle

\label{firstpage}

\begin{abstract}
The formation and evolution of supermassive black holes (SMBHs) remains an open question in the field of modern cosmology. The detection of nanohertz (n-Hz) gravitational waves via pulsar timing arrays (PTAs) in the form of individual events and the stochastic gravitational wave background (SGWB) offers a promising avenue for studying SMBH evolution across cosmic time, with SGWB signal being the immediately detectable signal with the currently accessible telescope sensitivities. 
By connecting the galaxy properties in the large scale (Gpc scale) cosmological simulation such as \texttt{MICECAT} with the small scale ($\sim$ Mpc scale) galaxy simulations from \texttt{ROMULUS}, we show that different scenarios of galaxy--SMBH evolution with redshift leads to a frequency-dependent spatial anisotropy in the SGWB signal. The presence of slow evolution of the SMBHs in the Universe leads to a pronounced blue anisotropic spectrum of the SGWB. In contrast, if SMBHs grow faster in the Universe in lighter galaxies, the frequency-dependent spatial anisotropy exhibits a more flattened anisotropic spectrum. This additional aspect of the SGWB signal on top of the monopole SGWB signal, can give insight on how the SMBHs form in the high redshift Universe and its interplay with the galaxy formation from future measurements.

\end{abstract}

\begin{keywords} 
gravitational waves, black hole mergers, cosmology: miscellaneous
\end{keywords}

\section{Introduction}
The groundbreaking observation of gravitational waves (GW) by the LIGO-Virgo-KAGRA (LVK) collaboration \citep{abbott2016observation} has opened up new possibilities for understanding the complex nature of our Universe. The LVK collaboration has been discovering compact binary mergers, which include binary black holes (BBHs), binary neutron stars (BNSs), and neutron star-black holes (NSBHs) mergers, since 2015. \citep{abbott2021gwtc}. Pulsar Timing Array (PTA) \citep{verbiest2022pulsar,manchester2013international,burke2019astrophysics} operating in the $10^{-9}-10^{-7}$ Hz range promises to complement these findings and provide an additional avenue for studying the undiscovered domains of the Universe. The PTA consortia, which include the North American Observatory for Gravitational Waves (NANOGrav; \cite{mclaughlin2013north}), European PTA (EPTA; \cite{desvignes2016high}), Parkes PTA (PPTA; \cite{manchester2013parkes}), Indian PTA (InPTA; \cite{joshi2018precision}), Chinese PTA (CPTA; \cite{xu2023searching}) and MeerKAT \citep{bailes2020meerkat}, search for the gravitational wave by accurately measuring the pulse arrival times of a set of pulsars. Each group observes a set of pulsars once every few weeks for periods of years. Recently, the PTA collaborations NANOGrav \citep{agazie2023nanograv}, EPTA+InPTA \citep{antoniadis2023second}, PPTA \citep{zic2023parkes}, and CPTA \citep{xu2023searching} have announced evidence of a stochastic gravitational wave background (SGWB).

The primary source of SGWB is believed to be the population of supermassive black hole binaries (SMBHBs) in the mass range $\sim$ $10^{7}-10^{10} M_{\odot}$ apart from cosmological sources \citep{burke2019astrophysics}. These SMBHs are believed to have formed from smaller seed BHs. The seed BHs grew by the hierarchical mergers and accretion of surrounding matter \citep{volonteri2007evolution,merloni2008synthesis,volonteri2009journey,jahnke2011non,ricarte2019tracing}. The merger history of SMBHs is highly dependent on the environment in which these BHs evolve. The environment surrounding the host galaxy influences both the growth of the SMBH and the host galaxy. This suggests the coevolution of both the galaxies and their central SMBHs. There is strong evidence for the correlation between the masses of SMBHs and their host galaxy properties from simulations and observational studies \citep{haring2004black,jahnke2011non,beifiori2012correlations,mcconnell2013revisiting,kormendy2013coevolution,reines2015relations,habouzit2021supermassive}. Therefore, to understand the evolutionary trajectories of both SMBHs and host galaxies, it is essential to conduct a study that involves the simultaneous examination of SMBHBs and their host galaxies. This can help us address various questions, such as the formation and evolution of SMBHBs, and types of galaxies favorable for hosting these binaries.

In this context, the study of the GWs from the SMBHBs can be very crucial. Future PTA experiments are expected to detect signals 
from individual low redshift SMBHBs. Apart from the individual sources at low redshifts, the SGWB from unresolved sources at higher redshifts can provide crucial insights into the population and evolution of SMBHBs across cosmic time. The shape and strength of the SGWB spectrum depend on the population of the SMBHBs and their environment. When two galaxies hosting SMBHs merge, an SMBHB is formed. These binaries lose energy to the environment through mechanisms such as dynamical friction, stellar loss-cone scattering, and viscous drag, in addition to gravitational wave (GW) radiation \citep{sampson2015constraining,kelley2017massive,chen2017efficient,izquierdo2022massive,bromley2023supermassive}. The evolution of SMBHBs driven solely by GW emission predicts a power-law GW background spectrum. However, the spectrum of the SGWB is expected to vary significantly depending on the environment in which these binaries evolve \citep{saeedzadeh2023shining}. Environmental effects become dominant at larger separations, i.e., when the binary emits at lower frequencies, thereby influencing the shape and steepness of the spectra. Consequently, the spectrum of the SGWB can provide valuable insights into the environment in which SMBHBs are formed. In addition to the isotropic SGWB, the SGWB is also expected to exhibit spatial anisotropy \citep{sato2024exploring,mingarelli2013characterizing,gardiner2024beyond}. The measurement of the anisotropy will provide complementary information on top of the isotropic SGWB.

In this work, we demonstrate how we can study the imprints of the evolution of SMBHs across redshift on the large angular scale anisotropies in the SGWB as a function of its spectrum. This figure shows that the study of the large angular anisotropies (about a degree-scale or larger; \cite{Janssen:2014dka,ali2021insights,pol2022forecasting,sato2024exploring}) as a function of GW frequencies which are accessible from the current and next generation radio antennas, can make it possible to explore the formation and evolution of the SMBHBs. One of the observable quantities from the n-Hz signal is the sky-averaged SGWB signal $\Omega_{\rm gw}(f)$ and its spatial fluctuations in the spherical harmonic space denoted by $C_{\ell}(f)$ at large scales (lower values of $\ell$).  However, one of the major challenges in predicting the spatial fluctuations of the signal at large scales arising from the mergers of the SMBHBs across cosmic redshifts based on cosmological simulations is the involvement of large dynamical length scales from the Gpc scale to sub-parsec scale. Therefore, in this work, we investigate the impact of the cosmic evolution of SMBHBs on the n-Hz SGWB signal by using a  \textit{large scale to small scale adaptive technique} as shown by a schematic diagram in Fig. \ref{Cross}. In this technique, we combine the results of simulations with analytical prescriptions across three length scales: Gpc scale, Mpc scale, and sub-pc scale. The detailed methodology of this technique will be discussed in Sec. \ref{Cross_Match}.

Throughout the paper, we make the simplistic assumption of circular orbits for the SMBHBs and disregard all other potential sources of n-Hz gravitational waves, apart from SMBHBs. The presence of eccentricity in SMBHBs has been studied in the past \citep{armitage2005eccentricity,iwasawa2011eccentric,bonetti2020eccentricity,gualandris2022eccentricity}. In this analysis, we primarily focus on the large-scale anisotropy in the n-Hz GW signal and its dependence on the cosmic evolution of SMBHs. Therefore, the impact of eccentricity is not important. The paper is organized as follows: in Sec. \ref{Formation}, we review the formation and evolution of SMBHs and SMBHBs; in Sec. \ref{Cross_Match}, we discuss the cross-matching of the small-scale effect to the large scale distribution of the galaxies; in Sec. \ref{GW_SMBH}, we briefly discuss the SGWB from SMBHB population; in Sec. \ref{Simulation}, we describe the simulation used to generate the SGWB signal. Then, in Sec. \ref{result} we discuss the results of the simulation. Finally, in Sec. \ref{conc}, we discuss the conclusions and future prospects.

\section{Formation and Evolution of SMBHB}\label{Formation}

In our current understanding, SMBHs are believed to form through the accretion of matter and the hierarchical merger of the initial seed BHs \citep{volonteri2003assembly,merloni2008synthesis,volonteri2010formation,woods2019revealing,chen2020dynamical,wirth2020formation,ni2022astrid}. The precise process leading to the formation of these seeds is still a subject of ongoing research, but they represent the foundational stage for SMBHs. The seed BHs in early halos grow by accreting halo gas and by hierarchical mergers. These BHs sink to the center of the halo due to dynamical friction and form the center of the galaxies \citep{volonteri2010formation,smith2019supermassive,woods2019revealing}. These galaxies further merge to give binary BHs. Although the fundamental mechanism that gives rise to the evolution of the SMBH may be known to some extent, we still do not comprehend many aspects and phenomena associated with the SMBH evolution. Some of the most important questions in this context are listed below.

\begin{enumerate}
    \item 
\textit{\textbf{What kind of galaxies host the SMBHBs?}: }

Several simulations, as well as observational analyses, indicate a correlation between the mass of SMBHs and the properties of their host galaxies, particularly bulge mass, stellar mass, luminosity, and velocity dispersion \citep{cattaneo2009role,haring2004black,beifiori2012correlations,kormendy2013coevolution,reines2015relations,mcconnell2013revisiting,muhamed2023mass}. This signifies the coevolution of SMBHs and their host galaxies. However, the interplay between the growth of SMBHs and galaxies is not yet well understood. Understanding the formation and evolution of both SMBHBs and their hosts requires not just their relationship at the local Universe but also its evolution with redshift \citep{ding2020mass,muhamed2023mass}.

The probability of a galaxy to host a SMBHB of a certain mass denoted as $P(M_{\rm BH},q|{ \rm \texttt{galaxy}})$, and its redshift evolution is crucial for addressing many questions regarding the SMBH population and its evolution. Here, $M_{\rm {1}}$ and $q$ represent the primary mass of the SMBHB and mass ratio, respectively, and '\texttt{galaxy}' denotes the properties of the host galaxies such as stellar mass, luminosity, velocity dispersion, etc. In this analysis, we use the stellar mass ($M_{*}$) of the host galaxy as an indicator to model the occupation of SMBHBs in galaxies.

\item \textit{\textbf{
How did SMBHs form and evolve over cosmic time?} }

The potential sources of the seed BHs include remnants of the first generation of stars, massive primordial black holes, and the collapse of massive stars formed in runaway collisions within dense stellar clusters \citep{latif2016formation,sassano2021light,volonteri2021origins}. These seed BHs co-evolve with their host galaxies accreting matter from their surrounding. Moreover, galaxy mergers, which were more common in the earlier, denser universe, contributed to the growth of SMBHs. Over billions of years, these processes allowed SMBHs to reach masses ranging from millions to billions of solar masses.

Studying the SMBH mass-galaxy stellar mass relation [$P(M_{\rm BH},q|M_{*},z$)] and its evolution can potentially provide insights into the intricacies of the growth mechanisms of SMBHs. This relation is thought to be influenced by various factors, including accretion rates, galaxy mergers, feedback processes, etc. By analyzing how this correlation changes over cosmic time, we can gain a deeper understanding of the formation and evolution of SMBHs.


\item \textit{\textbf{How efficiently SMBHBs merge in Hubble time for contributing to n-Hz GW signal?}} 

SMBHBs are the byproduct of galaxy mergers. These binaries spiral inward under the influence of their surroundings, such as dynamical friction and viscous drag. These processes are known to be effective until binary separation is greater than a parsec. However, the GW radiation becomes an effective source of orbital energy decay when the binary separation is much below parsec (around $10^{-3}-10^{-2}$ parsec). The challenge of how these binaries can come close enough to merge within the Hubble time is known as the final parsec problem \citep{milosavljevic2003final,vasiliev2015final,koo2023final}. For the binaries to emit GWs in the PTA-sensitive range and eventually merge, the environmental processes mentioned above must be effective enough to bring the binaries to the separation where the GW radiation becomes an efficient process of energy loss.

The efficiency of environmental hardening directly influences the observed distribution of binary separation, $P(a|M_{*})$ as well as the occupation fraction of the binary host galaxy, where '$a$' is the semi-major axis of the binary. A distribution with relatively more sources emitting at higher frequencies could indicate that binaries are efficiently being transported to smaller separations. $P(a|M_{*})$ can be further factored as the coalescence rate of the binaries at a given separation ($dN_a/dt_c$) and time spent by a binary at that separation ($dt_c/da$). The quantity $dt_c/da$ depends on the environment condition in which binary evolves. The effect of the environment on $dt_c/da$ can be modeled as \citep{sampson2015constraining,saeedzadeh2023shining}
\begin{equation}
    \frac{\tau_{\rm h}}{\tau_{\rm GW}}(f_{r}) =  \Big[1 + \beta \Big(\frac{f_{r}}{f_{t}}\Big)^{-\kappa}\Big]^{-\gamma},
\label{tau}
\end{equation}
where $\tau_h(f_{\rm r})$ and ${\tau_{\rm GW}}(f_r)$ represent the binary hardening time and the GW hardening time of the binary respectively, $f_t$ represents the transition frequency above which the GW emission becomes dominant over the environmental effect. $\kappa$, $\gamma$, and $\beta$ are parameters that depend on the environmental properties of the host galaxy like stellar mass, gas density, stellar density, velocity dispersion, etc. Here, $\kappa$ governs the spectral dependence of the environmental effects, $\gamma$ controls the overall shape of the environmental effects, and $\beta$ governs the overall strength of the environmental effect in comparison to the hardening effect solely due to gravitational waves. We refer to \cite{saeedzadeh2023shining} for a detailed description of the parameters and their dependence on the galactic properties.

\end{enumerate}

\section{Adaptive technique: Cosmological scales to sub-pc scale}\label{Cross_Match}

\begin{figure*}
    \centering
    \includegraphics[width=\textwidth]{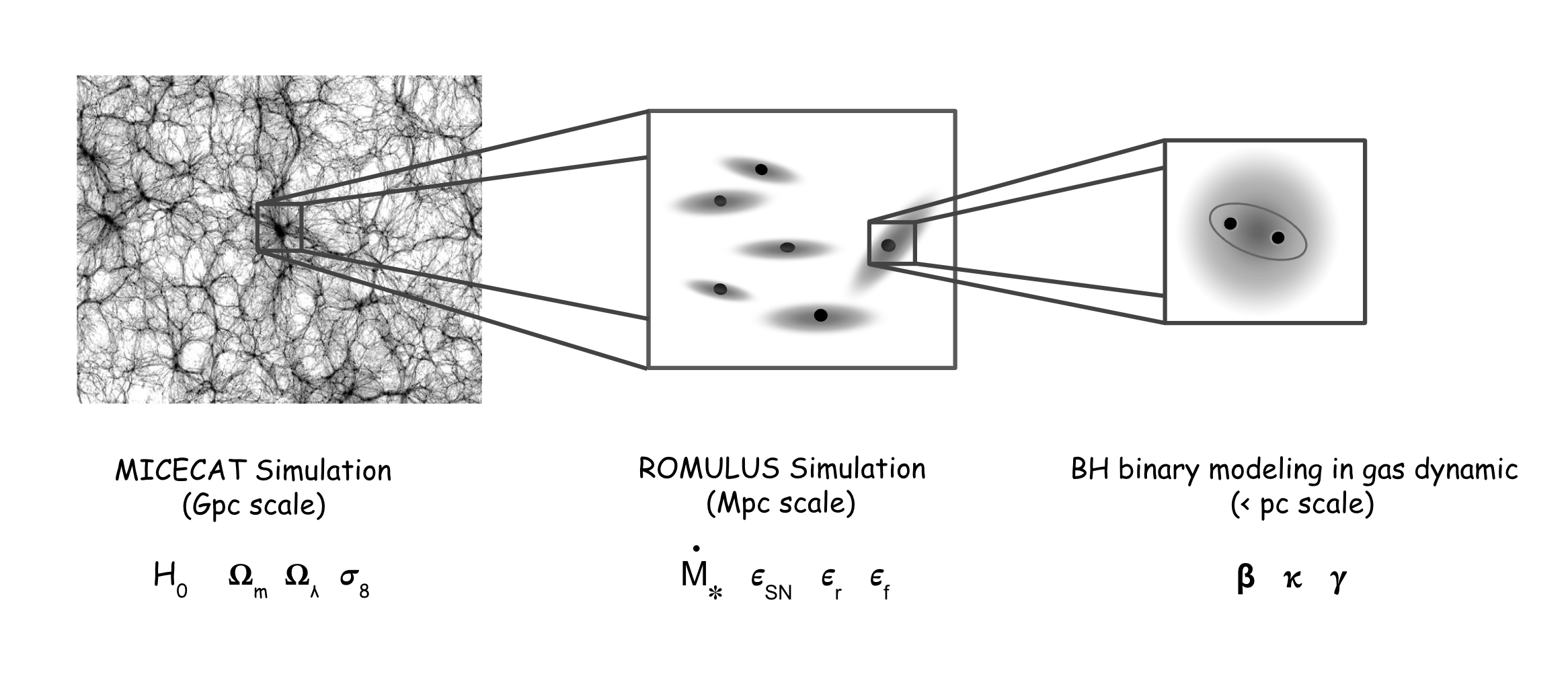}
  
    \caption{A schematic diagram illustrating simulations of galaxy and SMBH evolution at different scales.}
    \label{Cross}
\end{figure*}

Cosmological simulations, which aim to model the evolution of the large-scale distribution of the Universe, are among the most memory-intensive simulations in astrophysics. Simulating a $\rm Gpc^{3}$ comoving volume at kpc resolution can be very time-consuming, making kpc-resolution simulations feasible only for small volumes. High resolution simulation captures detailed halo and galaxy dynamics, while large-scale simulations offer insights into the large-scale structure of the Universe. The small-scale dynamics do not affect the inference of the large-scale simulation due to the low resolution of these simulations. Therefore, it becomes essential to bridge the large-scale cosmological simulation to the small-scale simulations by modeling the impact of the physics at the small scale on the larger-scale structure for studying the signatures of the cosmic evolution of SMBHBs from the n-Hz SGWB signal. A schematic diagram explaining the adaptive technique is shown in Fig. \ref{Cross}.

One such large-scale cosmological simulation is the MICE Grand Challenge (\texttt{MICE-GC}) \citep{fosalba2015mice,crocce2015mice}. \texttt{MICE-GC} is a dark matter-only simulation with halo mass resolution of around $10^{11} M_{\odot}$. \texttt{MICECAT} is a halo and galaxy catalog derived from the \texttt{MICE-GC} simulation. The galaxies are assigned to the halos by halo occupation distribution \citep{scoccimarro2001many,berlind2002halo,crocce2015mice,wechsler2018connection} and abundance matching technique \citep{vale2004linking,conroy2006modeling,crocce2015mice,wechsler2018connection,springel2018first}. The catalog provides quantities such as stellar mass and star formation rate of galaxies, which are crucial for bridging different-scale physics \citep{saeedzadeh2023shining}. In the following subsections, we briefly delve into cross-matching the dynamics at different scales in cosmological simulations. The application of this technique is not only limited to  \texttt{MICECAT} but can be extended to other large cosmological simulations. We will extend this to other large-scale cosmological simulations in future work.

\subsection{Cosmological scales to galaxy scales}\label{cos_gal}
Within the domain of cosmological simulations, bridging the vast range of scales from the largest cosmological structures to individual galaxies presents a significant challenge. Simulations such as \texttt{MICECAT} and \texttt{ROMULUS} \citep{tremmel2017romulus,butsky2019ultraviolet,tremmel2020formation,saeedzadeh2023cool} are at the forefront of this effort. \texttt{MICECAT} covers one-eighth of the sky covering around 4 $\rm Gpc^{3}$ of comoving volume. It focuses on the evolution of large-scale structures, tracing the growth of cosmic web-like filaments and the formation of dark matter particles over billions of years. On the other hand, the \texttt{ROMULUS} simulations are high-resolution, cosmological simulations that explore the smaller, more intricate scales of the Universe. Among the \texttt{ROMULUS} simulations, \texttt{ROMULUS25} is the leading simulation that extends over a uniform volume of $\rm 25 ~ Mpc^{3}$ \citep{saeedzadeh2023shining,sharma2020black}. The simulation includes phenomena such as star formation, feedback mechanisms, and interactions with the surrounding environment. It employed a massively parallel tree+SPH code, CHaNGa \citep{menon2015adaptive,wadsley2017gasoline2}, and has the Plummer equivalent gravitational force softening of 250 pc, a maximum SPH resolution of 70 pc. This high resolution enables the tracking of the evolution of galaxies and the SMBHs to the sub-kpc scale. One of the key findings from the \texttt{ROMULUS25} simulation is the insights into the properties of the SMBH population and the intricate relationship between SMBH mass and the properties of their host galaxies and halos \citep{ricarte2021origins,saeedzadeh2023shining,tremmel2023enhanced}. These simulations reveal that SMBHBs crucial for PTA are more likely to reside in galaxies with high stellar and halo masses but low star formation rates (SFR) that reisdes at the center of galaxy group \citep{saeedzadeh2023shining}. Specifically, SMBHBs with chirp masses greater than $10^8 M_{\odot}$ are more prevalent in galaxies with stellar masses exceeding $10^{11} M_{\odot}$ \citep{saeedzadeh2023shining}. The characteristics of SMBHBs from a small-volume simulation of \texttt{ROMULUS25} can be complemented by cross-matching them with \texttt{MICECAT} galaxies. Using the knowledge gained from the \texttt{ROMULUS} simulations, we can effectively populate galaxies in \texttt{MICECAT} with SMBHBs using parametric relation between $M_{\rm BH}$ and $M_{*}$.

It is to be noted here that the \texttt{MICECAT} catalog used in this work is complete for DES-like surveys ($i<24$) out to z=1.4 \citep{dark2005dark}. Therefore, the completeness of the catalog is poor at high redshifts. As a result, the number of galaxies in the catalog at high redshifts is absent. This limits us from exploring high redshift (beyond $z>1.4$) SMBHB mergers from the simulated Universe. However, in standard scenarios, the mergers of SMBHs that can contribute to the n-Hz signal take a significant fraction of the age of the Universe, to come from kpc to sub-pc scale \citep{saeedzadeh2023shining}. This leads to a larger delay time from the time SMBHs form to the time SMBHBs can contribute to the n-Hz frequency band. As a result, the number of SMBHs that can contribute to the n-Hz signal at high redshift is lower, as also shown in \citet{saeedzadeh2023shining} from \texttt{ROMULUS25}.  From the simulation, we have seen that contributions to the n-Hz signal mostly arise from redshift $z<1.4$. Also, the masses of SMBHs are likely to be lighter at higher redshifts than at lower redshifts, as they have more cosmic time to gain masses through accretion, except for the scenario where massive BH seeds are formed by the direct collapse of massive gas clouds \citep{natarajan2023first}. Consequently, the masses that can contribute to the n-Hz signal at higher redshifts will be lighter than the sources at lower redshifts, making their contribution to the n-Hz signal weak. Due to both these effects, we expect the limitation of \texttt{MICECAT} due to the redshift cutoff not to be severe for our analysis.

The multiscale adaptive technique that we have developed in this analysis to make a simulation-based prediction of the spatial-spectral signal in the n-Hz frequency range for different parametric forms of SMBH mass and evolution can be applied to other cosmological simulations that are complete up to higher redshifts. This will be explored in future work.

\subsection{Galaxy scales to sub-pc scales}

\texttt{ROMULUS25} tracks the evolution of thousands of galaxies and SMBHs up to a gravitational softening length of 0.7 kpc. Below this
scale, the gravitational interactions between SMBHs and their surrounding environments play a significant role in the evolution of the SMBHBs. At these scales, the dynamics of SMBHBs are affected by environmental effects like dynamic friction, stellar loss cone scattering, and viscous drag. These dissipation processes decrease the residence time of the binary at different separations, thereby influencing the shape and slope of the SGWB spectrum. Eq. \ref{tau} represents the effect of the host galaxy environment on the SGWB spectrum, where we model this effect using three parameters: $\beta$, $\kappa$, and $\gamma$. A denser stellar and gaseous environment will result in a decrease in the environmental hardening timescale. This affects the distribution of orbital stages in the evolution of the SMBHBs. Modeling the sub-pc scales allows us to populate the host galaxies with the SMBHBs at different stages of their evolution.

\section{Gravitational Wave from SMBHBs} \label{GW_SMBH}
\begin{figure}
    \centering
    \includegraphics[width=8cm]{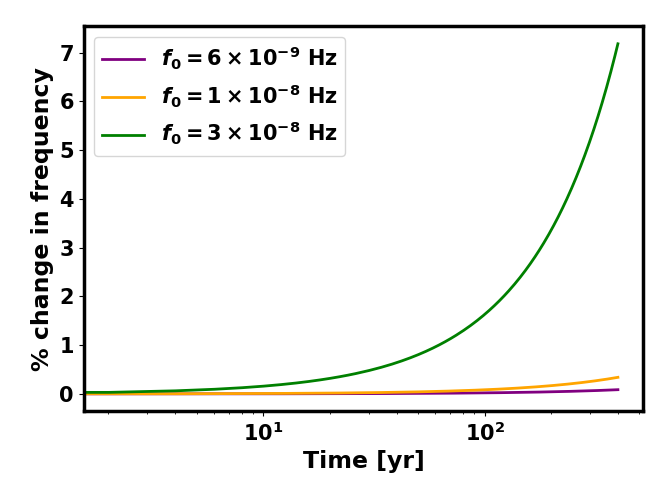}
    \caption{The change in frequency emitted by an SMBHB of chirp mass $10^{9} M_{\odot}$ when it starts with a frequency of $f_0= 6 \times 10^{-9}$ Hz, $f_0= 1 \times 10^{-8}$ Hz, and $f_0= 3 \times 10^{-8}$ Hz.}
    \label{freq_1}
\end{figure}

SMBHBs can emit GWs in the frequency range $10^{-9} - 10^{-7}$ Hz detectable by PTA. The orbit of the SMBHBs in this frequency range is highly stable.
Fig. \ref{freq_1} illustrates the evolution of the frequency of GW emitted by SMBHBs starting with three different frequencies. In particular, SMBHBs with a chirp mass $10^{9} M_{\odot}$ emitting at a frequency as high as $3 \times 10^{-8}$ Hz demonstrate a minimal frequency deviation over the observational timescale of PTAs. This makes SMBHBs a fairly monochromatic source of GW.

The SGWB is the superposition of all the GW sources that are unresolved as an individual source. It is defined as the GW energy density per unit logarithmic frequency divided by the critical energy density of the Universe ($\rho_c c^2$). The SGWB density due to the sources in a unit solid angle in the direction $\hat{n}$ can be expressed as a function of frequency $f$  as \citep{phinney2001practical,sesana2008stochastic,christensen2018stochastic}
\begin{equation}
    \begin{aligned}
   \Omega_{\rm gw}(f, \hat n) = \frac{1}{\rho_c c^2} & \int \prod_{i}^{n} d\theta_i \int_{z_{\rm min}}^{\infty} dz \frac{d\tilde{V}}{dz} \\
    &\times \left[\frac{1}{1+z} \frac{d^{n+4}N(z,\Theta_n,\hat n)}{d\Theta_n dVdt_r}\right] \left[\frac{1+z}{4 \pi d_{L}^{2} c}\right]  \\
    &\times \left[ f_r \frac{dE_{\rm{gw}}(f,\Theta_n, \hat n)}{df_r} \right],
    \end{aligned}
    \label{SGWB}
\end{equation}

where $V$ represents the comoving volume, $dV$ =  $d\tilde{V} d\omega$, with $d\omega$ representing the solid angle element in the direction $\hat{n}$. The term $\frac{ d^{n+4}N(z,\Theta_n,\hat n)}{ d\Theta_n dVdt_r}$ is the mergers rate of the binaries per unit comoving volume at redshift z, at the direction $\hat n$, and per unit source properties, $\frac{ dE_{\rm{gw}}(f,\Theta_n,\hat n)}{df_r}$ is the energy emitted by the source per unit source frame frequency ($\rm f_r$) and $\Theta_n$ = $\{\theta_i\}_{i=1}^{n}$ denotes the GW source parameters. The quantity $\frac{ d^{n+4}N}{ d\Theta_n dVdt_r}$ can be expressed as

\begin{equation}
    \frac{ d^{n+4}N(z,\Theta_n,\hat n)}{ d\Theta_n dV dt_r} = \frac{ d^{n+4}N(z,\Theta_n,\hat n)}{d\Theta_n dV d\ln f_r} \times \frac{ d\ln f_r}{ dt_r},
    \label{Nf}
\end{equation}
where $\frac{ d^{n+4}N(z,\Theta_n,\hat n)}{ d\Theta_n dV d\ln f_r}$ represents the comoving number density of the sources emitting in logarithmic frequency interval. Substituting eq. \ref{Nf} into eq. \ref{SGWB}, gives
\begin{equation}
    \begin{aligned}
     \Omega_{\rm gw}(f, \hat n) = \frac{1}{\rho_c c^2} & \int  \prod_{i}^{n} d\theta_i \int\limits_{z_{ min}}^{\infty} ~dz~ \frac{ d\tilde{V}}{ dz} \\
    &\times \bigg[\frac{ d^{n+4}N(z,\Theta_n,\hat n)}{ d\Theta_n dVd\ln f_r}\bigg]  \bigg[\frac{1}{4 \pi d_{L}^{2} c}\bigg]  \\
    &\times \bigg[\frac{d\ln f_r}{\rm dt_r} \frac{ dE_{\rm{gw}}(f,\Theta_n, \hat n)}{{ d\ln f_r}} \bigg].
    \end{aligned}
    \label{SGWB2}
\end{equation}
The term  $\frac{ d\ln f_r}{ dt_r} \frac{ dE_{\rm{gw}}(\Theta_n, \hat n)}{d\ln f_r}$ is the flux of energy emitted by a binary emitting at frequency $ f_{r}$ with
\begin{equation}
    \frac{dE_{\rm gw}}{d\ln f_r} = \frac{(G \pi)^{2/3}  M_c^{5/3} \times f_{r}^{2/3}}{3},
\end{equation}
where $M_c$ is the chirp mass of the SMBHB. The relative fluctuation in $\Omega_{\rm gw}(f,\hat{n})$ can be defined as

\begin{equation}
    \Delta\Omega_{\rm gw}(f,\hat{n}) \equiv {\frac{\Omega_{\rm gw}(f,\hat{n}) - \overline{\Omega}_{\rm gw}(f)}{\overline{\Omega}_{\rm gw}(f)}},
\end{equation}
where $\Delta\Omega_{\rm gw}(f,\hat{n})$ can be written in terms of spherical harmonics $Y_{\ell m}(\hat{n})$ as
\begin{equation}
    \Delta\Omega_{\rm gw}(\rm f,\hat{n}) = \sum\limits_{\ell} \sum\limits_{m = -\ell}^{\rm \ell} \omega_{\ell m}(f) Y_{\ell m}(\hat{n}). 
\end{equation}
The angular power spectrum ($C_\ell$) is given by
\begin{equation}
     C_{\ell}(f) = \frac{1}{(2\ell+1)}\sum\limits_{m} |\omega_{\ell m}(f)|^{2}.
\end{equation}
$C_{\ell}(f)$ is the angular power spectrum. It captures the angular fluctuation of the $\Delta\Omega_{\rm gw}(\rm f,\hat{n})$ at angular scale of $\Delta \theta = \pi/\ell$. The anisotropy depends on the higher order moment ($\ell > 0$). The relative fluctuations map $\Delta \Omega_{\rm gw}(f,\hat n)$ is a mean zero field and is directly related to the tracer of the galaxy density field in the Universe (as discussed later in the paper). The mean-zero fluctuations map $\Delta \Omega_{\rm gw}(f,\hat n)$ defined in this analysis is useful to recover an unbiased non-zero cross-correlation signal with the galaxy density field fluctuations map $\Delta \delta{\rm g}(\hat n)$, in contrast to $\Omega_{\rm gw}(f,\hat n)$ usually used in the literature \citep{NANOGrav:2023tcn, pol2022forecasting}.

\section{Simulation of the SGWB}\label{Simulation}

The strength and shape of the spectrum of the SGWB are influenced by several factors, including the mass distribution, number density, binary separation distribution, and redshift evolution of the population. A population with SMBHs of higher chirp mass will result in a stronger background. Similarly, the binary separation distribution will affect the overall shape of the spectra. The evolution of the population of SMBHBs with redshift depends on factors such as the merger history of galaxies, the delay time between galaxy mergers, and the formation of the binary, along with other factors like the accretion history of the SMBHs.

\subsection{SMBHB Population}

We aim to connect the SMBHB host galaxy properties to the population of SMBHB. 
The distribution of the source population in eq. \ref{SGWB2} can then be written with the quantities discussed in the Sec. \ref{Formation} as
\begin{equation}
    \begin{aligned}
     \frac{d^{7}N }{d\log( M_{*})d\log(M_{\rm BH})d q dV da} \propto& \frac{d^{4}N }{d\log( M_{*})dV}  P( M_{\rm BH},q, a| M_{*},z) \\
     =& \overbrace{\frac{d^{4}N }{d\log( M_{*}) dV}}^{MICECAT} \\
     &\overbrace{\times P( M_{\rm BH}| M_*,z) \times P(q| M_*,z)}^{ROMULUS}\\
     &\times  \overbrace{P(a| M_*,z)}^{sub-pc ~ physics},
    \end{aligned}
    \label{pop}
\end{equation}
where, the term $\frac{d^4N }{d\log( M_{*})dV}$ represents the number density of host galaxies with stellar mass, $M_{*}$, in the catalog. The terms $P( M_{\rm BH}| M_,z)$, $P(q| M_,z)$, and $P(a| M_*,z)$ represent the probability distributions of the primary mass ($M_{\rm BH}$), mass ratio ($q$), and binary separation ($a$), respectively, given the stellar mass of the host galaxy and its redshift. The $\frac{d^4N }{d\log( M_{*}) dV}$ can be directly obtained from the \texttt{MICECAT} simulation. The distributions $P( M_{\rm BH}| M_,z)$ and $P(q| M_,z)$ can be estimated from the \texttt{ROMULUS} simulation \citep{saeedzadeh2023shining} and by using parametric forms discussed below, the distribution $P(a| M_*,z)$ can be modeled by including several dissipative processes that are active at sub-pc scales. Although the SMBH mass can also depend on several other galactic properties, for simplicity, we consider only the stellar mass dependence of the SMBHB masses. We model $P( M_{\rm BH}| M_,z)$ and $P(q| M_,z)$ as
\begin{equation}
    P( M_{\rm BH}|  M_{*},z) \propto \mathcal{N}( \mathrm{Log}_{10}(M_{\rm BH})| \mathrm{Log}_{10}(M_{\mu}),\sigma_m),
    \label{M1}
\end{equation}

\begin{equation}
     P(q|  M_{*},z) \propto \bigg\{
    \begin{array}{cl}
    & 1/q, \quad  0.01 < q < 1,\\
    & 0, ~~ else ,
    \end{array}
    \label{q}
\end{equation}

where $\mathcal{N}$ is the normalised Gaussian with standard deviation  $\sigma_m$ and mean $ \mathrm{Log}_{10}(M_{\mu})$, which is modeled as
\begin{equation}
\mathrm{Log}_{10}(M_{\mu})= \eta + \rho ~ \mathrm{Log}_{10}( M_{*}/10^{11} M_{\odot}) + \nu  z. 
\label{Mmu2}
\end{equation}

where $\eta$, $\rho$, and $\nu$ are free parameters, where $\nu$ controls the redshift evolution of the relation. This parametric form will make it feasible to obtain the theoretical signal of $\Omega_{\rm gw}(f)$ and $C_{\ell}(f)$ for a vast range of astrophysical models efficiently, which can be incorporated in a Bayesian framework to constrain from n-Hz data in future. 

\begin{figure}
  \subfigure[]{\label{Mass_Dist}
    \centering
    \includegraphics[width=\linewidth,trim={0.cm 0  0 0.cm},clip]{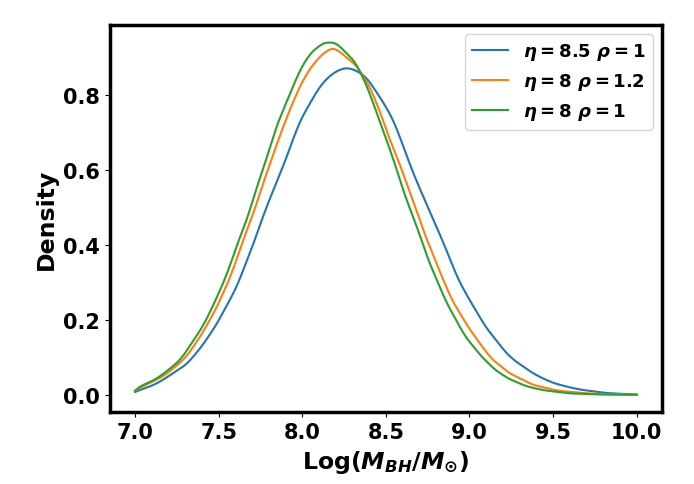}}
  \subfigure[]{\label{Mass_Dist_z}
    \centering
    \includegraphics[width=\linewidth,trim={0.cm 0cm  0cm 0.cm},clip]{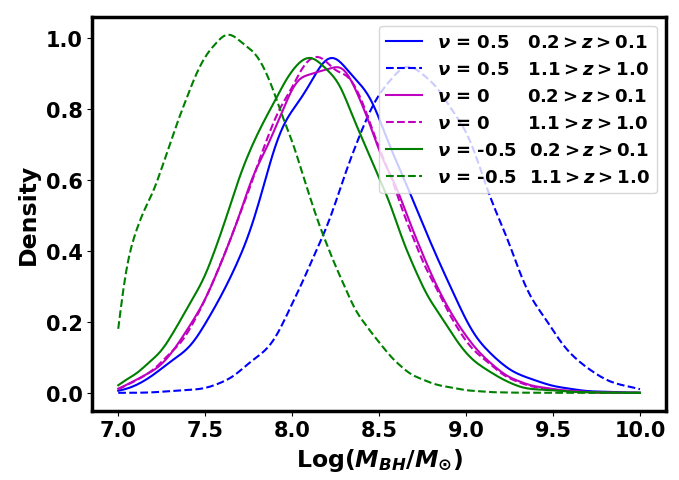}}
  \caption{(a) Mass distribution of the primary mass ($M_{\rm BH}$) of SMBHBs for different $\eta$ and $\rho$. (b) Mass distribution of the primary mass ($M_{\rm BH}$) of SMBHBs for different $\nu$ at different redshift bins. }
  \label{Mass}
\end{figure}

In Fig. \ref{Mass_Dist}, we show the primary mass ($M_{\rm BH}$) distribution of the simulated SMBHBs using the above equations for different values of $\eta$ and $\rho$. The scenario with higher values of $\eta$ and $\rho$ favors higher mass BHs, as shown in the figure. In Fig. \ref{Mass_Dist_z}, we display the primary mass distribution of the SMBHBs for different $\nu$ values at various redshift bins. A negative (or positive) value of $\nu$ indicates a preference for higher mass at low (or high) redshifts. At low redshifts, the distributions for all values of $\nu$ considered here appear to be very similar. However, at high redshifts, the distribution varies significantly for different $\nu$ values. This allows us to see the redshift evolution of SMBH masses in the Universe. From standard cosmological scenarios, we expect the heavier SMBHs to be more likely in the low redshifts, which is captured by $\nu<0$.  

\begin{figure*}
    \centering
    \begin{minipage}{0.4\linewidth}
        \centering
        \subfigure[]{\label{M1_Ms_1}
        \includegraphics[width=\linewidth]{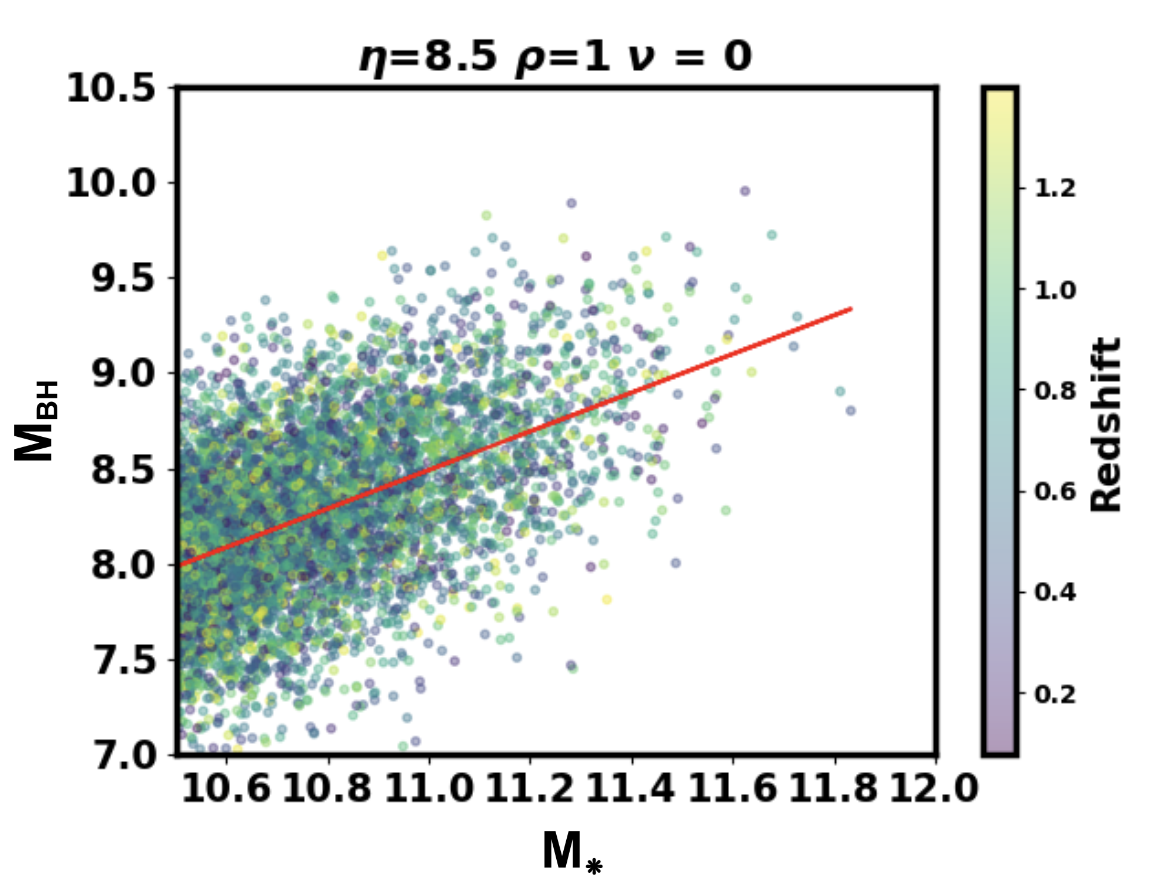}}
    \end{minipage}%
    \begin{minipage}{0.4\linewidth}
        \centering
        \subfigure[]{\label{M1_Ms_2}
        \includegraphics[width=\linewidth]{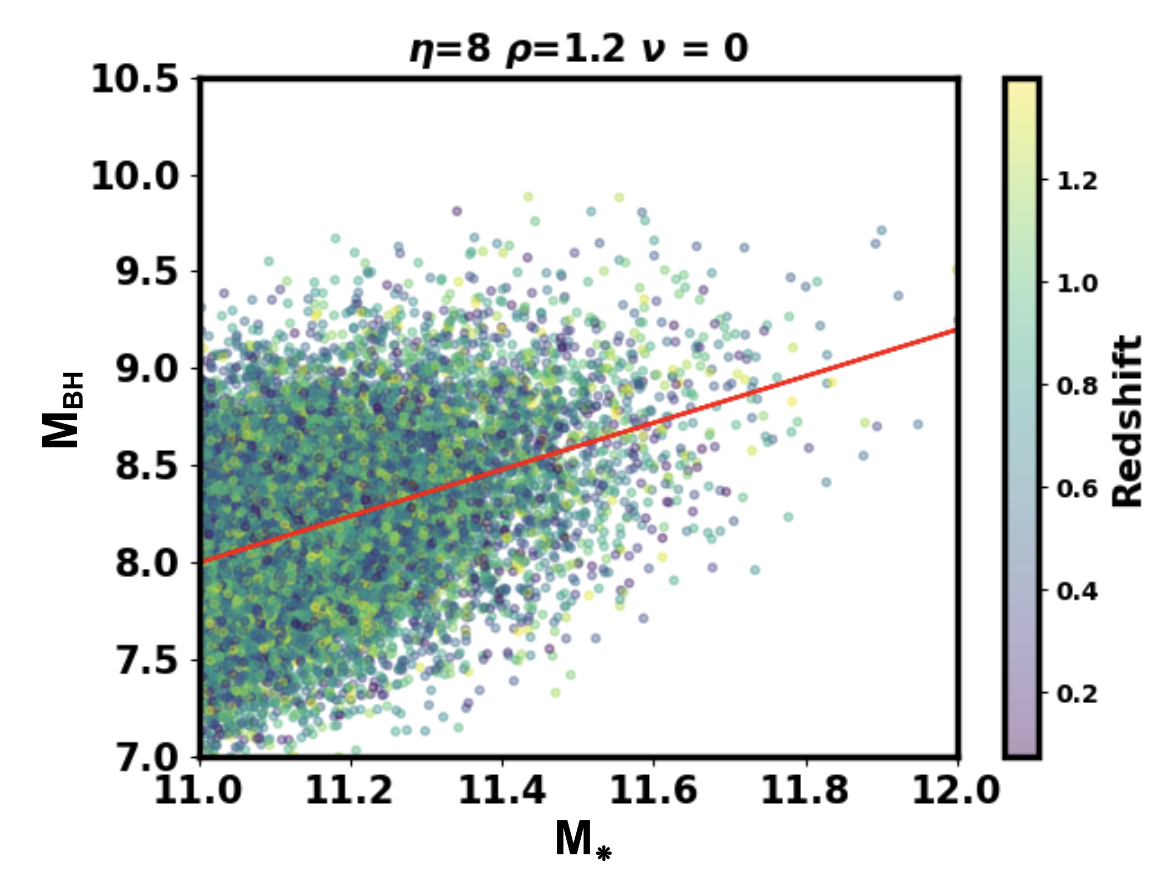}}
    \end{minipage}%
    
    \begin{minipage}{0.4\linewidth}
        \centering
        \subfigure[]{\label{M1_Ms_3}
        \includegraphics[width=\linewidth]{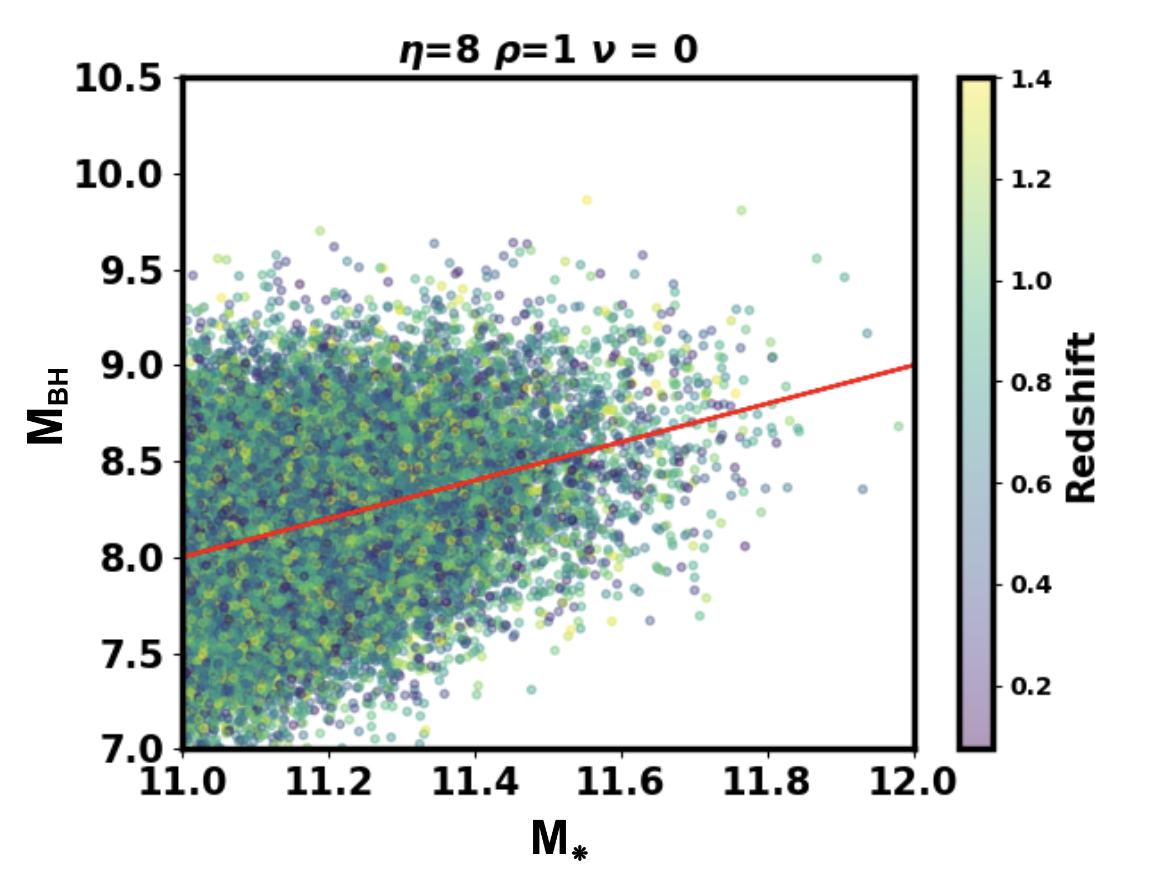}}
    \end{minipage}
    \begin{minipage}{0.4\linewidth}
        \centering
        \subfigure[]{\label{M1_Ms_4}
        \includegraphics[width=\linewidth]{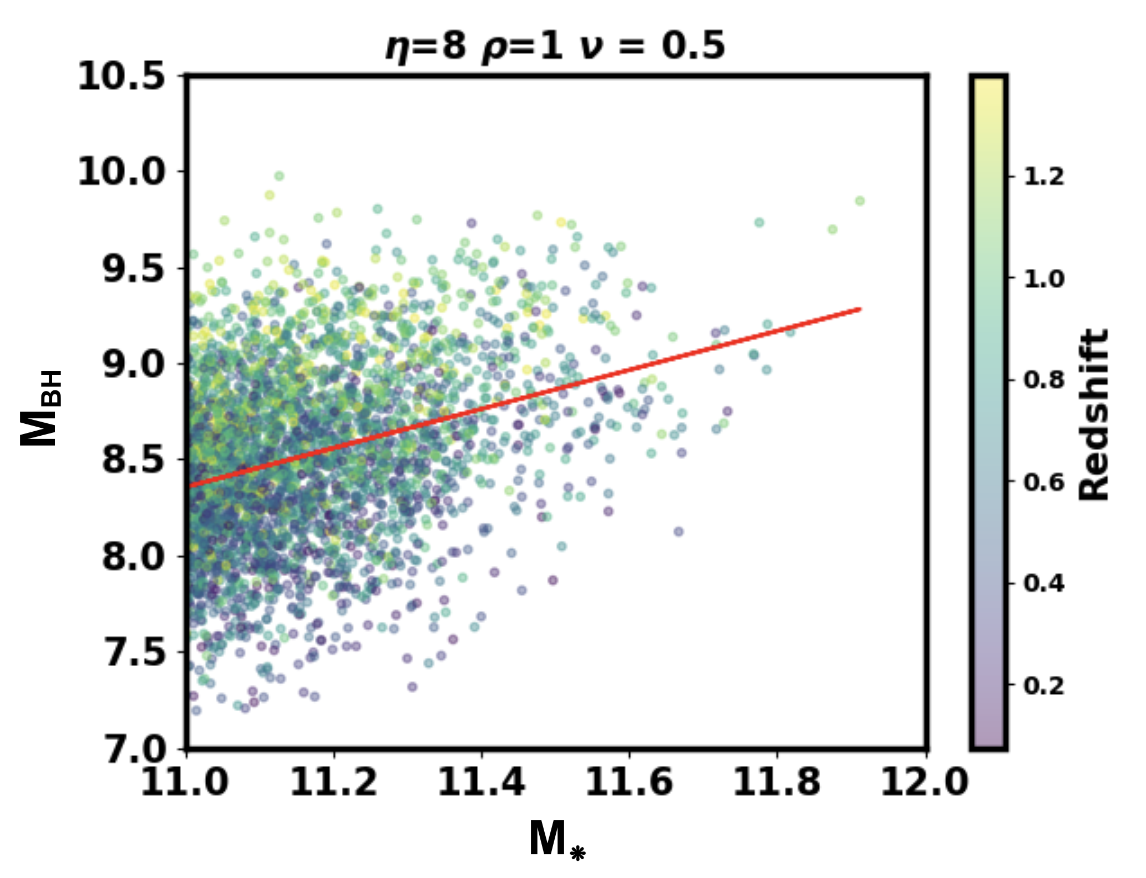}}
    \end{minipage}

    \begin{minipage}{0.4\linewidth}
        \centering
        \subfigure[]{\label{M1_Ms_5}
        \includegraphics[width=\linewidth]{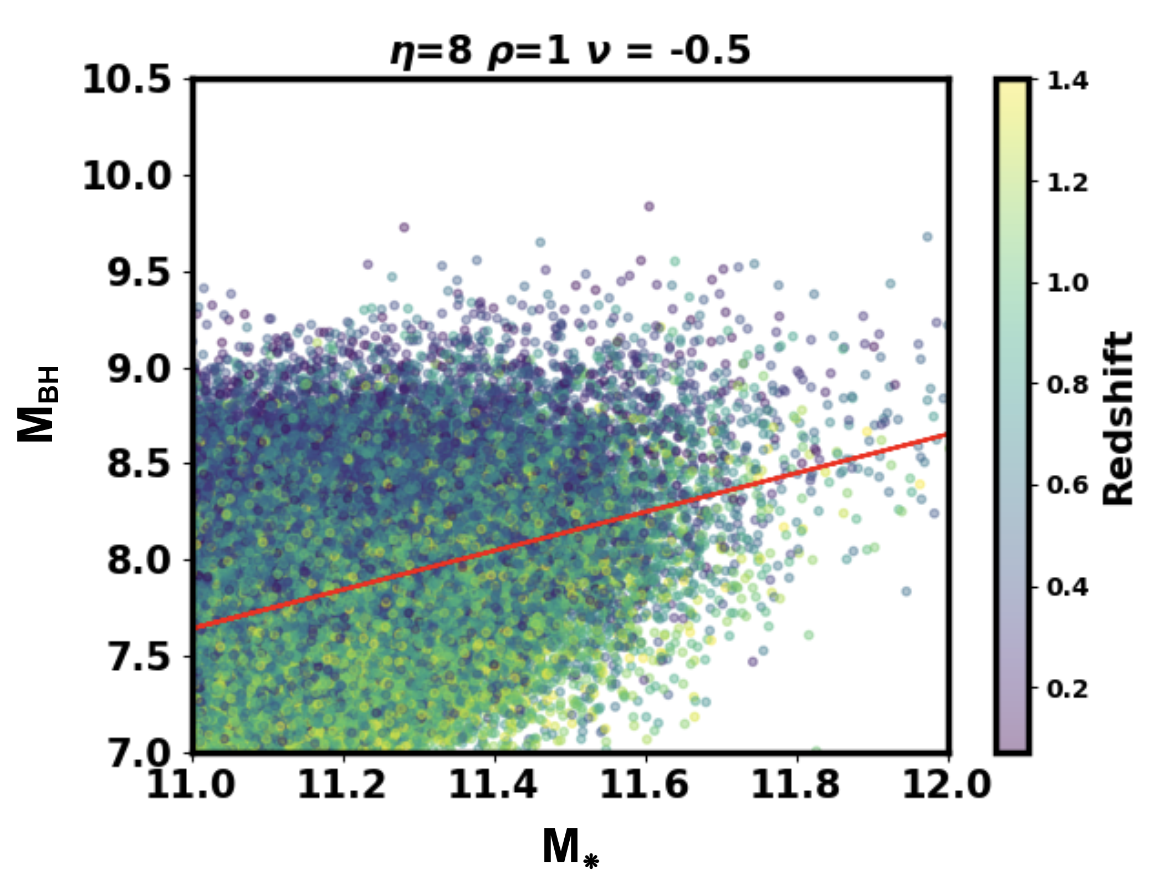}}
    \end{minipage}
    \caption{Plots showing the primary mass ($M_{\rm BH}$) of the SMBHBs and the stellar mass ($M_{*}$) of the host galaxies from the simulations, for different values of $\eta$ and $\rho$, and $\nu$, with their redshift represented by colormap}
    \label{M1_Ms}    
\end{figure*}

In Fig. \ref{M1_Ms}, we show the primary mass of the SMBHs ($M_{\rm BH}$) and the stellar mass ($M_{*}$) of the host galaxy of the simulations, with their redshift represented in color-bar for different values of $\eta$ and $\rho$, and $\nu$. In Fig. \ref{M1_Ms_1}, \ref{M1_Ms_2}, and \ref{M1_Ms_3}, $\nu$= 0, therefore the SMBH mass-Stellar mass ($M_{\rm BH} - M_{*}$) relation does not evolve with the redshift. In Fig. \ref{M1_Ms_4} and \ref{M1_Ms_5}, $\nu$= 0.5 and $\nu$= -0.5 respectively. The value $\nu=0.5$ indicates that there are heavier BHs at higher redshifts, while $\nu=-0.5$ indicates that there are lighter BHs at higher redshifts. This trend can be observed in the two figures, where most of the low-redshift BHs lie below and above the fitted lines in Fig. \ref{M1_Ms_4} and \ref{M1_Ms_5}, respectively. If heavier SMBHs are more prevalent at lower redshifts, it would suggest that SMBHs are continuously growing even at low redshifts. However, if there is a population of SMBHs with heavier BHs at higher redshifts than at lower redshifts, this could imply that the growth of heavier SMBHs ceased at lower redshifts.  This parameter $\nu$ also has a significant impact on the SGWB anisotropy signal as we will discuss in the results section.

One of the next important ingredients for simulating the SGWB signal is to model the probability distribution of the orbital separation of SMBHs in galaxies of different stellar masses across cosmic redshift, $P( a| M_{*})$. This can be written in terms of the coalescing rate and residence time of a source at a particular orbital separation by
\begin{equation}
    P( a|  M_{*}) \propto \frac{dN_a}{dt_c} \times \frac{dt_c}{da}, 
    \label{freq_dist}
\end{equation}
where $\frac{dN_a}{dt_c}$ and $\frac{dt_c}{da}$ represent the coalescence rate and the residence time of the binaries at the separation of '$a$', respectively. We have assumed a power-law form of $\frac{dN_a}{dt_c}$, where $\delta$ is a power law index and $\mu$ controls its redshift evolution, given by:
\begin{equation}
    \frac{\mathrm{d}N_a}{\mathrm{d}t_c} \propto a^{\delta*(z-z_0)^{\mu}}.
    \label{Flow_eq}
\end{equation}
The binary separation '$a$' corresponds directly to the emitted frequency ($f_r$) of the binary, which is an observable quantity of gravitational waves in terms of redshifted frequency $f_z= f_r/(1+z)$ for a source contributing from redshift $z$. Thus, we prefer to model  the separation distribution $P(a|M_*)$ with the probability distribution of frequencies emitted by the binary  $P(f_r|M_*)$ as 
\begin{equation}
    \begin{aligned}
        P( f_r|  M_{*}) = & ~ P( a|  M_{*}) \times \frac{da}{df_r}\\
        \propto &  ~ \frac{dN_f}{dt_c} \times \frac{dt_c}{df_r}, 
    \label{freq_dist}
    \end{aligned}
\end{equation}
where $\frac{da}{df_r}$ is the Jacobian  of the transformation from '$a$' to '$f_r$'. We can write the coalescence rate as a function of frequency as: 
\begin{equation}
    \begin{aligned}
        \frac{\mathrm{d}N_f}{\mathrm{d}t_c} 
        \propto & ~ f_r^{-\frac{2}{3}\delta*(z-z_0)^{\mu}}\\
        \propto & ~ f_r^{\delta_{f}*(z-z_0)^{\mu}}
    \label{Flow_eq}
    \end{aligned}
\end{equation}
where the parameters $\delta_f$, $z_0$, and $\mu$ control the frequency dependence and the redshift dependence of the coalescing rate across cosmic time. The residence time of a binary at a frequency $f_r$, $\frac{\mathrm{d}t_c}{\mathrm{d}f_r}$, is modeled as \citep{saeedzadeh2023shining}
\begin{equation}
    \frac{\mathrm{d}t_c}{\mathrm{d}f_r} \propto f_{r}^{-11/3} \times \frac{\tau_h}{\tau_{\rm GW}}(f_r).
    \label{pop}
\end{equation}
As discussed in the Sec. \ref{Formation}, the quantity  $\frac{\tau_h}{\tau_{\rm GW}}(f_r)$ represents the environment hardening time scale and the factor $f_{r}^{-11/3}$ captures the GW only hardening time scale. The residence time of the binary due to only GW emission at frequency $f_r$ is proportional to $f_{r}^{-11/3}$. The factor also represents the frequency distribution of the SMBHB population in the absence of the environmental effect and the frequency-dependent coalescence rate.

\begin{figure*}
    \centering
    \begin{minipage}{0.4\linewidth}
        \centering
        \subfigure[]{\label{N5}
        \includegraphics[width=\linewidth]{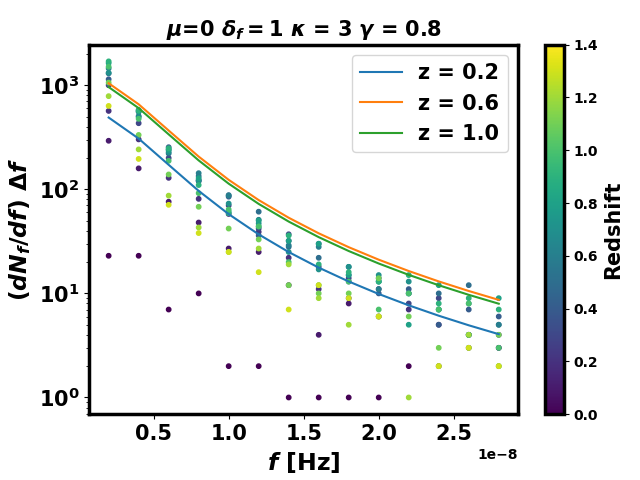}}
    \end{minipage}
    \begin{minipage}{0.4\linewidth}
        \centering
        \subfigure[]{\label{N2}
        \includegraphics[width=\linewidth]{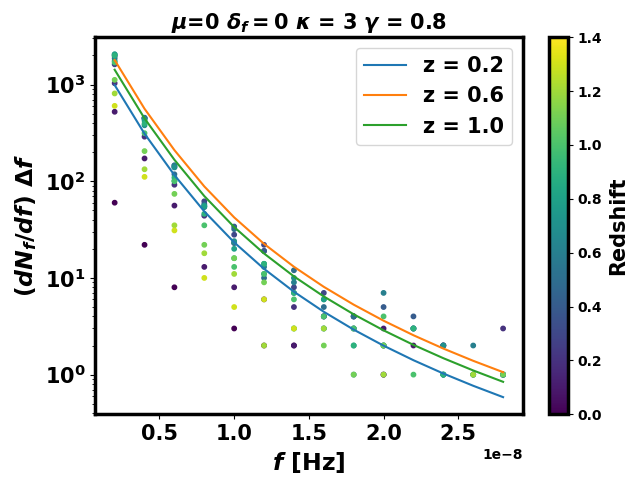}}
    \end{minipage}

    \begin{minipage}{0.4\linewidth}
        \centering
        \subfigure[]{\label{N4}
        \includegraphics[width=\linewidth]{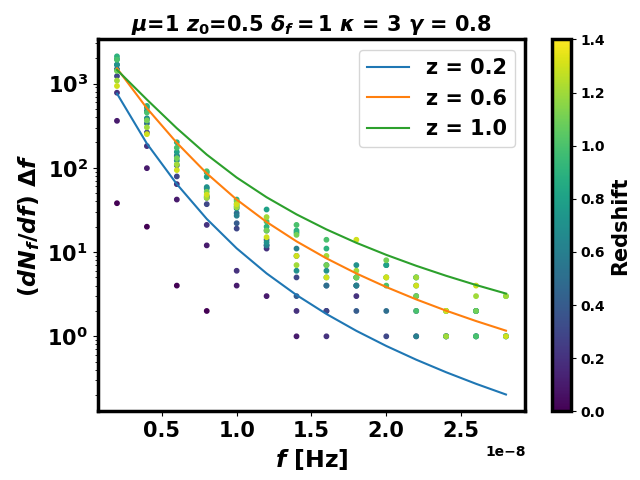}}
    \end{minipage}
    \begin{minipage}{0.4\linewidth}
        \centering
        \subfigure[]{\label{N3}
        \includegraphics[width=\linewidth]{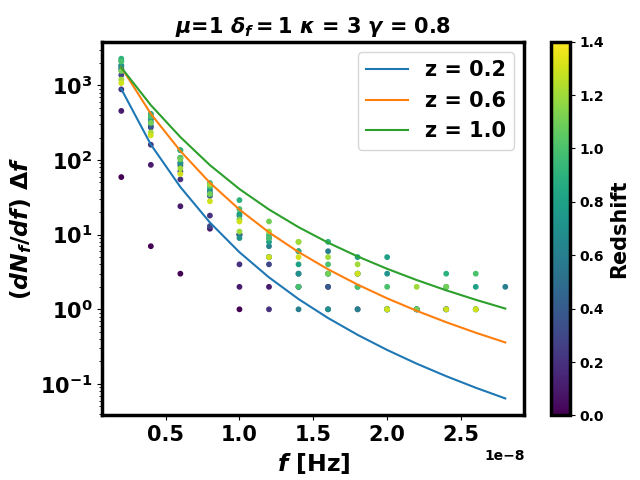}}
    \end{minipage}
    \caption{Plots showing the number of sources emitting in each frequency bin $\Delta f$ = $2\times 10^{-9}$ Hz, and redshift bin $\Delta z$ = 0.1,  as a function of the observer frame Frequency ($f$) for different $\mu$, $z_0$, $\delta_f$, and $\kappa$. The redshift is represented by a colormap.}
    \label{dNdf}    
\end{figure*}

In Fig. \ref{dNdf}, we illustrate the $\frac{dN_f}{df}$ vs $f$, of the sources in the simulation along with the redshift represented by colormap as well as theoretical curves at three different redshifts, for different values of $\mu$ and $z_0$. In Fig. \ref{N5} and Fig. \ref{N2} we represent the case where $\mu$ = 0 and hence no redshift dependence of the frequency distribution, as can be seen from the analytical curves. In Fig. \ref{N4} and Fig. \ref{N3}, we show the same quantity for the $\mu$ = 1, with $z_0$ = 0.5 and $z_0$ = 1, respectively. Here, the slope of the curve decreases with the redshift (less negative). This means that there are comparatively more sources at low frequencies than at high frequencies at low redshifts. In these cases, the number of sources emitting at high frequency is even smaller than that of redshift independent case. Consequently, the SGWB power spectrum is expected to be less steep toward higher frequencies.  Also, since the number of sources at high frequency is smaller, the anisotropy is expected to be larger in these cases. 

The analytical curves show the increasing trend in the slope with decreasing redshift. In this case, the slope of the curve is expected to be even larger than in the last case at low frequencies. The distribution of the sources at lower frequencies fits the analytical curve better than at the higher frequencies due to the larger number of sources at lower frequencies than higher frequencies. As the number of SMBHs must be an integer and cannot be fractional, there is a discrepancy between the analytical and simulated distributions of the sources at higher frequencies, where the analytical distribution predicts $(dN_f/df) \Delta f < 1$. Due to the Poisson fluctuation in the number of sources, the scatter of the sources at high frequencies is higher than at the low frequencies.

\subsection{Simulation Methodology}

We have constructed all the parts required to calculate the larger-scale observation of the GW background signal from the sub-pc scale SMBHBs. The sub-pc scale physics provides us with the number distribution of binary separations ($a$) of the sources, which determines the occupation fraction of galaxies with those separations. The \texttt{ROMULUS} simulation informs us about the masses of the SMBHBs that reside in galaxies with a given stellar mass. The \texttt{MICECAT} catalog provides the number density of galaxies with a given stellar mass. The relationship between the SMBH mass and the galaxy stellar mass, combined with the galaxy catalog, allows us to determine the population of SMBHBs and, consequently, its large-scale distribution. The anisotropic SGWB can be determined by calculating the GW signal from individual sources. Combining all these terms, we can express the term $ \frac{d^{n+4}N(\hat n)}{d\Theta_n dV df_r}$ in Eq. \ref{SGWB2} as Eq. \ref{pop}

We sample the frequency of the binary using eq. \ref{freq_dist} in the range accessible from the PTA detector. The minimum frequency as well as the size of the frequency bin ($\Delta$ f) of the PTA is given by the total observation time (T) of pulsar timing, which is 1/T. The maximum frequency ($\rm f_{max}$) is given by  Nyquist frequency (1/(2 $\Delta$t)), where $\Delta$t is the interval between two consecutive observations. For our analysis, we assume $\Delta$ f = $2\times 10^{-9}$ Hz.

\begin{figure}
  \subfigure[]{\label{Omega_eta}
    \centering
    \includegraphics[width=\linewidth,trim={0.cm 0  0 0.cm},clip]{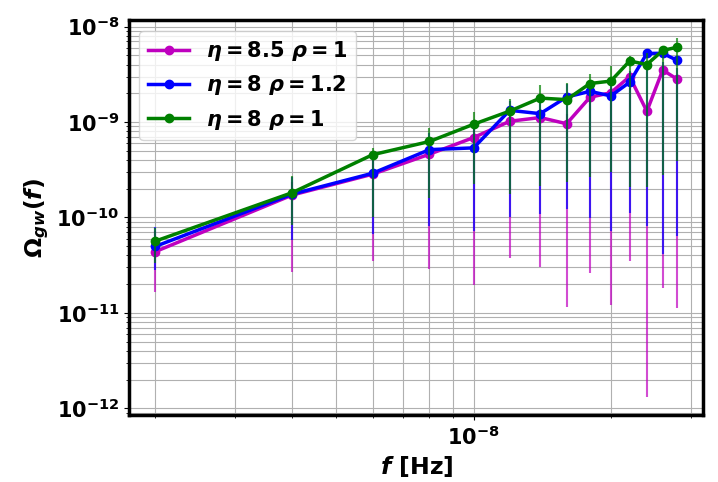}}
  \subfigure[]{\label{Cl_eta}
    \centering
    \includegraphics[width=\linewidth,trim={0.cm 0cm  0cm 0.cm},clip]{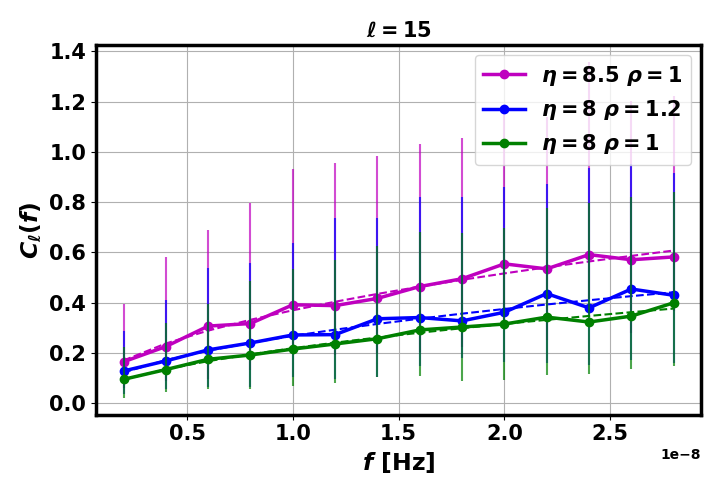}}
   \caption{SMBH mass-Stellar mass : (a) $\Omega_{\rm gw}$(f) for different values of $\eta$ and $\rho$ (b) $C_l(f)$ vs Frequency ($f$) for different values of $\eta$ and $\rho$, for $\ell$ = 15.}
  \label{eta}
\end{figure}

\begin{figure}
  \subfigure[]{\label{Omega_nu}
    \centering
    \includegraphics[width=\linewidth,trim={0.cm 0  0 0.cm},clip]{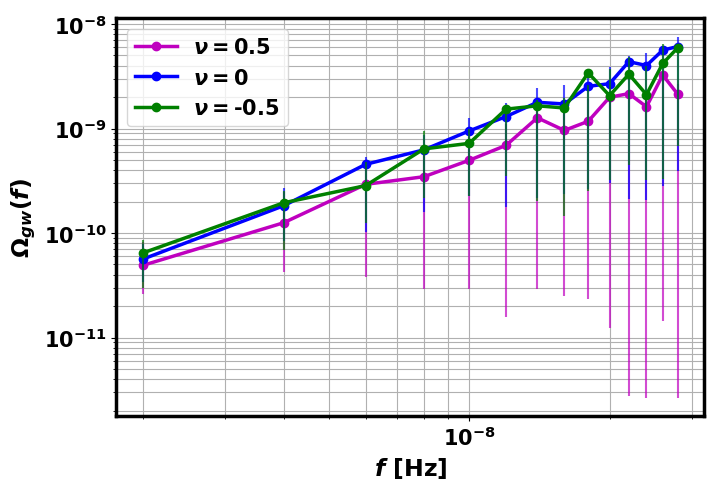}}
  \subfigure[]{\label{Cl_nu}
    \centering
    \includegraphics[width=\linewidth,trim={0.cm 0cm  0cm 0.cm},clip]{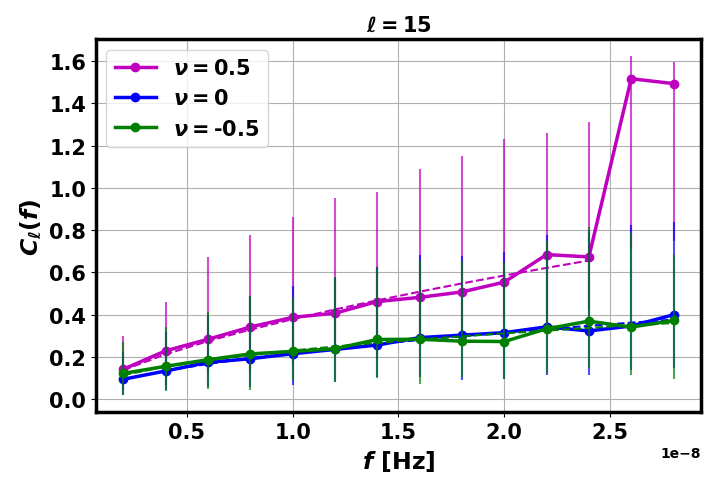}}
  \caption{SMBH mass-Stellar mass evolution : (a) $\Omega_{\rm gw}$(f) for different values of $\nu$ (b) $C_l(f)$ vs Frequency ($f$) for different values of  $\nu$, for $\ell$ = 15.}
  \label{nu}
\end{figure}

\begin{figure}
  \subfigure[]{\label{Omega_env}
    \centering
    \includegraphics[width=\linewidth,trim={0.cm 0  0 0.cm},clip]{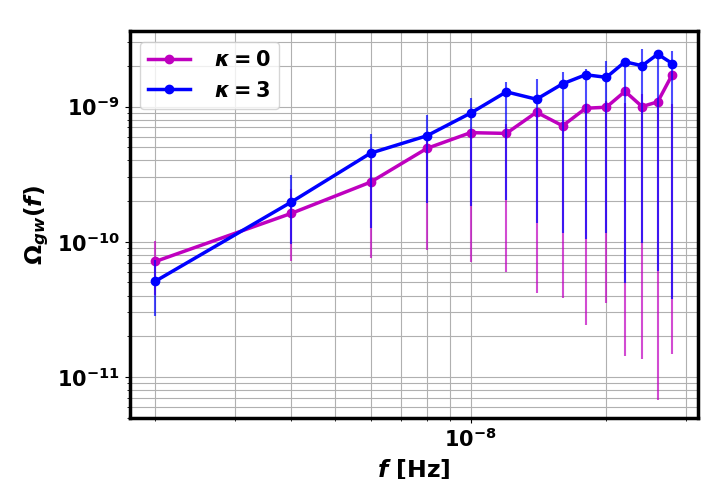}}
  \subfigure[]{\label{Cl_env}
    \centering
    \includegraphics[width=\linewidth,trim={0.cm 0cm  0cm 0.cm},clip]{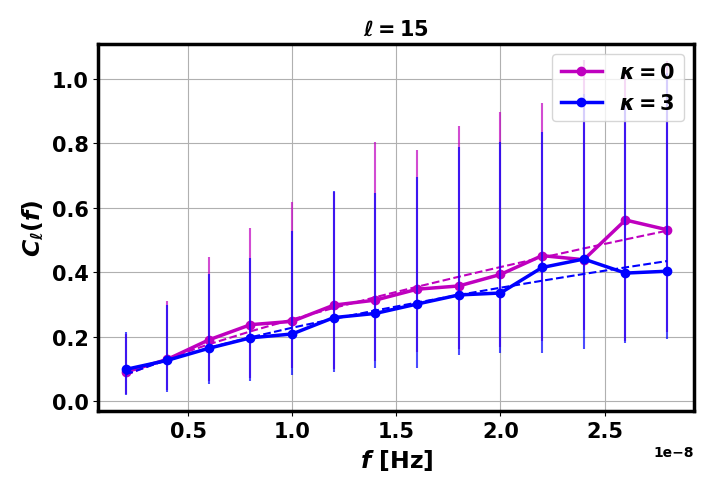}}
  \caption{Environmental effect : (a) $\Omega_{\rm gw}$(f) for different values of $\kappa$ (b) $C_l(f)$ vs Frequency ($f$) for different values of $\kappa$, for $\ell$ = 15.}
  \label{env}
\end{figure}

\begin{figure}
  \subfigure[]{\label{Omega_z0}
    \centering
    \includegraphics[width=\linewidth,trim={0.cm 0  0 0.cm},clip]{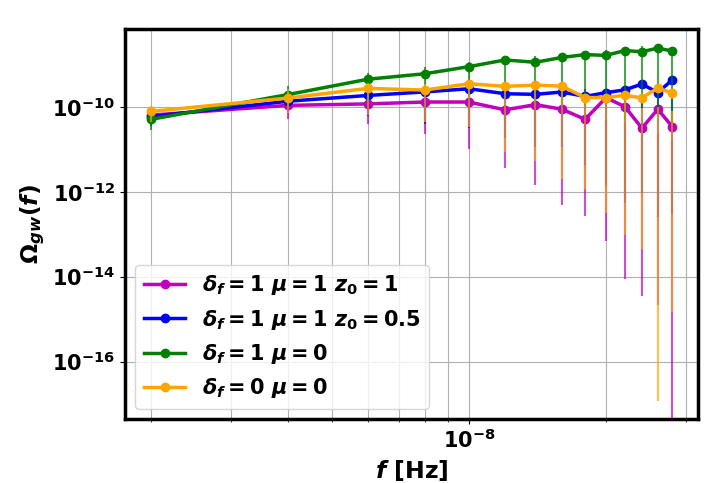}}
  \subfigure[]{\label{Cl_z0}
    \centering
    \includegraphics[width=\linewidth,trim={0.cm 0cm  0cm 0.cm},clip]{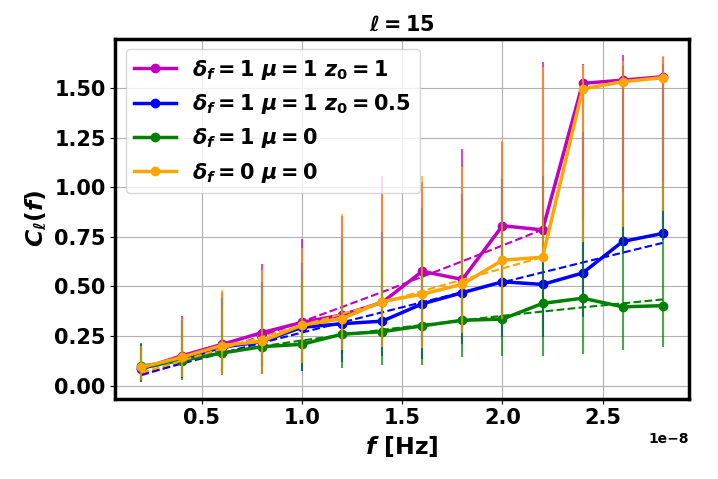}}
  \caption{Coalescence rate : (a) $\Omega_{\rm gw}$(f) for different values of $\delta_f$, $\mu$ and $z_0$ (b) $C_l(f)$ vs Frequency ($f$) for different values of $\delta_f$, $\mu$ and $z_0$, for $\ell$ = 15.}
  \label{z0}
\end{figure}

 We calculate the SGWB by Monte Carlo sampling of the population using eq. \ref{pop} and adding the contribution from each sampled source using eq. \ref{SGWB2}. We first assign the component masses of the binary to the host galaxies. For this, we use \texttt{MICECAT} galaxy catalog.  We simulate the background for different SMBH mass-galaxy stellar mass (SMBH-Stellar mass) relations and different host galaxy environmental scenarios. Simulations by \citep{saeedzadeh2023shining} have shown that most of the SMBHBs important for the PTA band reside in galaxies with high stellar mass and low star formation rate (SFR). The SMBHBs with mass   $> 10^{8} M_{\odot}$ are more likely to be found in the galaxies with stellar mass above $10^{11} M_{\odot}$ \citep{saeedzadeh2023shining}. We only select the host galaxies of \texttt{MICECAT} with stellar mass ($\rm M_{*}$) greater than that corresponding to the $\rm M_{\mu} = 10^{8} M_{\odot}$ in eq. \ref{Mmu2}.

 We then sample a primary BH mass ($\rm M_{\rm BH}$) using eq. \ref{M1} and the mass ratio q using eq. \ref{q} and obtain different Monte Carlo realizations of the SGWB corresponding to different realizations of SMBHBs distribution. If the SGWB signal is driven by a large number of sources (as in the lower frequencies, see Fig. \ref{dNdf}), then the distribution is Gaussian, and the use of mode, median, and mean represents the underlying distribution. However, for higher frequencies, we have fewer sources (see Fig. \ref{dNdf}), resulting in a non-Gaussian distribution of the SGWB signal. In this case, the mode is a better representative of the underlying distribution than the mean or median. In order to obtain the total number of SMBHBs occupying the galaxies, we normalize the realizations of the SGWB such that the mode of the distribution agrees with the data of the n-Hz signal observed by PTA. As a mode of distribution captures the most likely value of the underlying population (even for a non-Gaussian distribution), we expect the observed realization of SGWB of the Universe to be a typical distribution for a given underlying population model of the GW source. So, we have chosen the mode of the distribution as opposed to the mean or median in this analysis.

 We select sources from the sampled sources with equal probability, ensuring that the mode of the distribution of $\Omega_{\rm gw}$(f) in each frequency bin agrees with the current estimate from the NANOGrav 15-year data release \citep{agazie2023nanograv,agazie2023nanogravSMBH}. The details of the fitting can be found in the Appendix. \ref{sec:appendixA}. In Table. \ref{table1}, we show different cases for the SMBH population models considered in this work along with the parameters of the fitted power law curve (B$\times(f/2 \times 10^{-9})^{\lambda} $ ) to the mode of the $C_{\ell}(f)$. In the first two cases (SMBH mass-Stellar mass and SMBH mass-Stellar mass evolution), we choose the number of sources contributing to the signal in each frequency bin such that it is comparable to the 1/8 th power law curve fitted to the 15-year data release of NANOGrav. The factor 1/8 is taken to match the 1/8 part of the sky spanned by the \texttt{MICECAT} catalog. In the third and fourth cases (Environmental effect and Coalescence rate), the quantity of interest is the frequency distribution, i.e., the shape of the $\Omega_{\rm gw}(f)$. Therefore, we fix the total number of binaries to be equal to that in the third scenario of the SMBH mass-Stellar mass case (which is obtained after normalizing to the PTA data), instead of normalizing the $\Omega_{\rm gw}(f)$ by NANOGrav data again.

\begin{table*}
  \centering
  \begin{tabular}{|c|c|c|c|c|c|c|c|c|c|c|c|c|}
    \hline
    \textbf{Case} &  \textbf{$\eta$} & \textbf{$\rho$} & \textbf{$\nu$} & $\mu$ & \textbf{$z_0$} & \textbf{$\delta_f$} & \textbf{$\kappa$} & \textbf{$\gamma$} & \textbf{$f_t$} & \textbf{$\sigma_m$} & \textbf{B ($\ell= 15$)} & \textbf{$\lambda$} \\
    \hline
      & 8.5 & 1  & 0 & 0 & -- & -- & -- & -- & 5 $\times 10^{-9}$ & 0.4 & 0.17 & 0.48\\

    {$M_{\rm BH}$ and  $M_*$ relation}  $\qquad\qquad\qquad\qquad$  & 8 & 1.2  & 0  & 0 & -- & -- & -- & -- & 5 $\times 10^{-9}$ & 0.4  & 0.122 & 0.49  \\

     & 8 & 1  & 0 & 0 & -- & -- & -- & --  & 5 $\times 10^{-9}$ & 0.4 & 0.094 & 0.53 \\
    \hline
     & 8 & 1 & 0.5 & 0 & -- & -- & -- & --   & 5 $\times 10^{-9}$ & 0.4 &  0.1375 & 0.62 \\
    
    {Redshift evolution of $M_{\rm BH}$ and $M_*$ relation}  & 8 & 1 & 0 & 0  & -- & -- & -- & --  & 5 $\times 10^{-9}$ & 0.4 & 0.094 & 0.53 \\
    
       & 8 & 1 & -0.5 & 0  & -- & -- & -- & --  & 5 $\times 10^{-9}$ & 0.4 & 0.115 & 0.43\\
    \hline

    {Environmental effect} $\quad\qquad\qquad \qquad \qquad$  & 8 & 1 & 0 & 0 & -- & 1 & 0 & 0.8  & 5 $\times 10^{-9}$ & 0.4 & 0.0813 & 0.71\\

     & 8 & 1 & 0 & 0 & -- & 1 & 3 & 0.8 & 5 $\times 10^{-9}$ & 0.4  & 0.66 & 0.0825 \\
    \hline
    
       & 8 & 1 & 0 & 1 & 1 & 1 & 3 & 0.8 & 5 $\times 10^{-9}$ & 0.4 & 0.0525 & 1.13 \\

    {Coalescence rate} $\quad\qquad\qquad\qquad\qquad \quad $  & 8 & 1 & 0 & 1 & 0.5 & 1 & 3 & 0.8  & 5 $\times 10^{-9}$ & 0.4 &  0.059 & 0.94\\
       & 8 & 1 & 0 & 0 & -- & 1 & 3 & 0.8 & 5 $\times 10^{-9}$ & 0.4 & 0.0825 & 0.63\\
       & 8 & 1 & 0 & 0 & -- & 0 & 3 & 0.8  & 5 $\times 10^{-9}$ & 0.4 & 0.065 & 0.96\\

    \hline
    
    \end{tabular}
    \caption{Table representing all the cases of SMBHB population considered in the work. The presence of dash for parameters \textbf{$\delta$}, \textbf{$\kappa$} and \textbf{$\gamma$} in certain cases implies that the mode of the $\Omega_{\rm gw}$ is normalized by 50 \% confidence level of the power law curve parameters (A and $\alpha$) fitted to the 15-year data release of NANOGrav. The value of the fitted parameter B and $\lambda$ of a power law to the mode of the $C_{\ell}(f)$ (over 1000 realizations) vs $f$ is also shown.}
\label{table1}
\end{table*}

\section{Results: Isotropic and Anisotropic SGWB from simulations}\label{result}
Quantifying the n-Hz SGWB signal using both monopole power spectrum $\Omega_{\rm gw}(f)$ and angular power spectrum $C_{\ell}(f)$ as a function of frequencies brings complementary information.  
The isotropic SGWB signal $\Omega_{\rm gw}(f)$ measures the all-sky-averaged signal as a function of frequency, representing the net contribution across the observable Universe. However, the amplitude of the signal cannot distinguish between the scenarios on whether the signal is arising from a few very massive SMBHBs present at low redshift, or a large number of moderately massive SMBHBs spanning from very high redshift. Also, it cannot explore the underlying correlation with the galaxy properties. On the other hand, spatial anisotropy in SGWB measured in terms of $C_{\ell}(f)$ captures the spatial distribution of the SMBHBs. If the SMBHs form efficiently in the high redshift Universe and merge efficiently, the number of sources will be higher, and the spatial anisotropy will be less pronounced. In contrast, if the SMBH formation is not efficient at high redshifts, and they also evolve and merge inefficiently, then the spatial anisotropy is expected to be more significant. Moreover, as the number distribution of galaxies in heavier halos is higher at low redshift than at the higher redshifts if the host galaxies of the SMBHs are arising from heavier halos than the lighter halos, then the corresponding redshift distribution of the SMBHs will be more at low redshift. This will also exhibit a different spatial distribution of the SMBHBs and hence a different spatial distribution of the n-Hz signal. We describe below a few cases explored in this work to show the impact of SMBHB evolution on the spatial-spectral anisotropy of the SGWB signal. Moreover, since the number distribution of galaxies in heavier halos is higher at low redshifts compared to higher redshifts, if the host galaxies of the SMBHBs occupy heavier halos rather than lighter ones, then SMBHBs will predominantly occur in lower redshifts.

\textbf{SMBH mass and Stellar mass relation}: 
The varied scaling relation of the SMBH mass with the host galaxy properties will result in varying levels of anisotropy. The SGWB signal from two different SMBH mass functions with similar power spectra will result in different levels of anisotropy. This is because the population with larger mass BHs will require a relatively smaller number of binaries to achieve the same power spectrum. This becomes a very essential factor in breaking the degeneracy between the mass function and the number of sources. 

In Fig. \ref{Omega_eta}, we show the mode of $ \Omega_{\rm gw}(f)$ (over 1000 realizations) vs frequency for different scenarios, along with a 68 \% confidence interval around the mode, for different SMBH mass-Stellar mass relations. We present the mode of the $\Omega_{\rm gw}(f)$ as well as $C_{\ell}(f)$  instead of the mean, as the distributions of both quantities are highly skewed. \texttt{MICECAT} only covers one-eighth of the sky, so the magnitude of $\Omega_{\rm gw}(f)$ is one-eighth that of actual all sky $\Omega_{\rm gw}(f)$. The parameter $\eta$ captures the minimum mass of the galaxies needed to host the SMBHBs that are important for PTA sensitivity, while $\rho$ represents the slope of the relation. A curve characterized by high $\eta$ and $\rho$ values indicates a scenario in which galaxies host heavier SMBHs. The number of sources required to achieve a comparable SGWB density is small for these cases. As a consequence, the angular power spectrum ($C_\ell(f)$) for these cases is expected to be larger. This can be seen in Fig. \ref{Cl_eta} where we show the mode of the $C_\ell(f)$ versus frequency curves at $\ell$ = 15, along with a 68 \% confidence interval around the mode, for different values of parameters $\eta$ and $\rho$. It is important to note that the anisotropic signal is expected to be dominated by shot noise, and therefore, it is expected to remain similar at the scales that can be resolved by the PTA, as we will demonstrate in a later section. We therefore only show the result for single $\ell$. The  68 \% error bars on the $C_\ell(f)$  represent the uncertainty in the theoretical signal around the mode. This uncertainty arises because of cosmic variance. The cosmic variance results from the variation of the properties of individual sources in different realizations sampled from a given population of the SMBHB. In reality, we only have one realization of the universe. Consequently, the distribution of the properties of the source, such as the masses and number density of the source, introduces inherent uncertainty in the anisotropic signal $C_\ell$.



To understand the behavior of the anisotropy as a function of frequency, we infer the scaling of the anisotropy with frequency by fitting a power law $C_{\ell}(f)= B(\ell)\times$ $(f/2 \times 10^{-9})^{\lambda} $ to the mode of $ C_\ell(f)$. The value of parameters can be inferred from Table \ref{table1}. The power-law index for these cases is similar, with a slightly higher amplitude value for $\eta=8.5$ and $\rho=1$ compared to the other two cases. It is important to point out that, although all these cases have similar $\Omega_{\rm gw}(f)$, their $C_{\ell}(f)$ values differ across frequencies. The correlation between the masses of SMBHBs and their host galaxies provides valuable insight into the evolution and initial seed masses of SMBHs. In cases where $\eta$ is comparatively large, similar galaxies will host relatively larger SMBHs, suggesting that the SMBHs originated from large seed BHs or more efficient merger and accretion. This leads to a relatively larger anisotropy in the SGWB as discussed in the last paragraph. Similarly, a large value of $\rho$ suggests a stronger correlation of stellar mass with SMBH mass. A large value of $\rho$ will result in a higher anisotropy.

\textbf{Redshift evolution of SMBH mass and Stellar mass relation:} In Fig. \ref{nu}, we depict the mode of $ \Omega_{\rm gw}(f)$ and $C_\ell(f)$ (over 1000 realizations) vs. frequency, along with a confidence interval 68 \% around the mode for different scenarios of evolution of the $M_{\rm BH}-M_{*}$ relation with redshift, parameterized by $\nu$).  The anisotropy ($C_{\ell}(f)$) is larger for the case where $\nu$ is positive. However, the anisotropy is comparable for the cases of $\nu = 0 $ and $\nu = -0.5$. A positive value of $\nu$ represents a scenario where galaxies at higher redshifts host heavier SMBHBs, whereas a negative value of $\nu$ represents a scenario where galaxies at higher redshifts host lighter SMBHBs (see Fig. \ref{Mass_Dist} and Fig. \ref{Mass_Dist_z} for comparing their distribution). If heavier binaries occupy the center of a galaxy at high redshifts, a smaller number of binaries is required to generate the same SGWB. The anisotropy for cases with a negative value of $\nu$ will naturally be lower, as a larger number of binaries are required to generate the same SGWB compared to cases with a positive value of $\nu$, where we have heavier binaries at higher redshifts. The anisotropy is not very distinct between the cases when $\nu = 0$ and $\nu = -0.5$, despite there being a larger number of sources for $\nu = -0.5$. This is because lighter sources at higher redshifts in the case $\nu = -0.5$ contribute very little to the overall anisotropy. The power-law index ($\lambda$) and the amplitude (B) of the curve fitted to $C_{\ell}(f)$, tends to be slightly higher for positive values of $\nu$, as expected. 

It should be noted that there is a sharp jump in the $C_{\ell}(f)$ curve at $f = 2.6 \times 10^{-8}$ Hz for $\nu = 0.5$. This is because, at high-frequency bins, the expectation value of the number of sources in the low redshift bins is less than one (as can be seen for some cases in Fig. \ref{dNdf}). However, Monte Carlo sampling of the population mostly results in either one source or none, not a fraction of a source. Therefore, the sporadic sampling at high-frequency bins leads to a sudden increase in $C_{\ell}(f)$ at these frequencies.

 \begin{figure*}
  \centering
  \subfigure[]{\label{MT1}
    \includegraphics[width=0.45\linewidth,trim={0cm 0cm 0cm 0cm},clip]{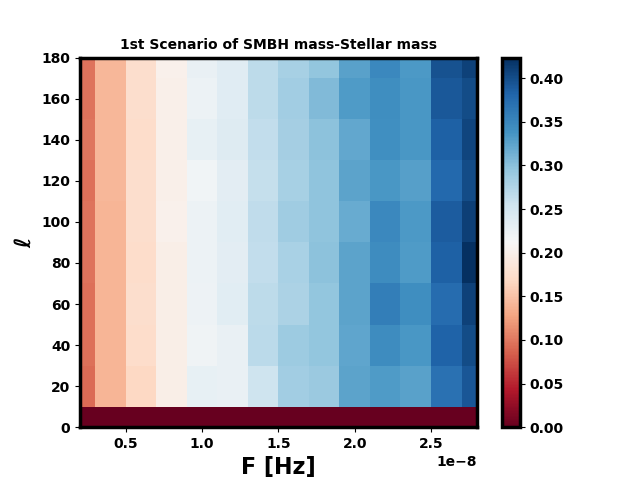}}
  \subfigure[]{\label{MT4}
    \includegraphics[width=0.45\linewidth,trim={0cm 0cm 0cm 0cm},clip]{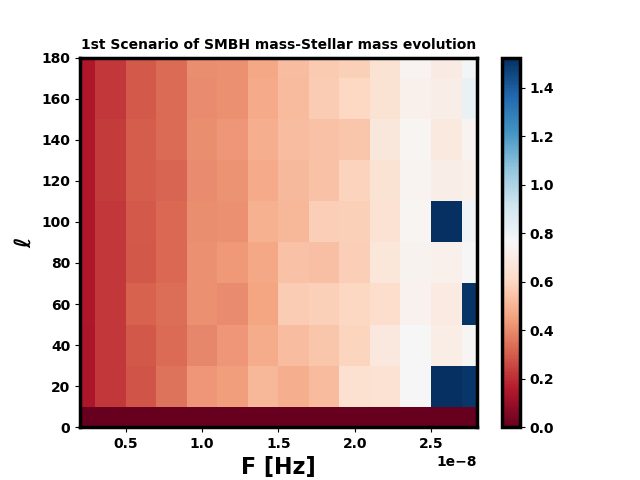}}
  
  \subfigure[]{\label{MT6}
    \includegraphics[width=0.45\linewidth,trim={0cm 0cm 0cm 0cm},clip]{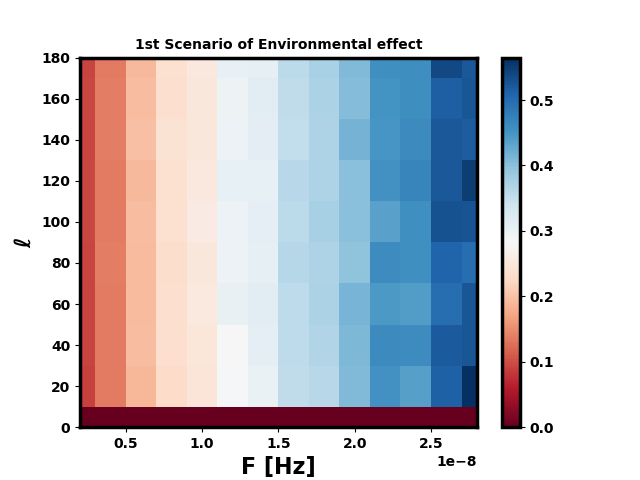}}
  \subfigure[]{\label{MT9}
    \includegraphics[width=0.45\linewidth,trim={0cm 0cm 0cm 0cm},clip]{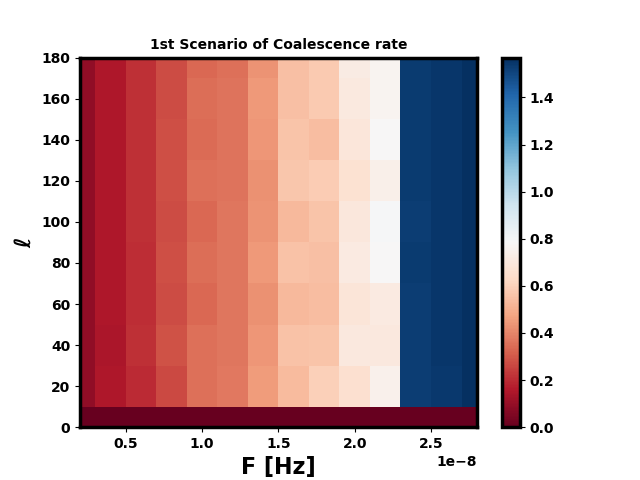}}
    
  \caption{Color matrix plot of the mode of $C_{\ell}(f)$ (over 1000 realizations) as a function of frequency ($f$) and multipole moment ($\ell$) for the first scenario of (a) the SMBH mass-Stellar mass case, (b) the SMBH mass-Stellar mass evolution case, (c) the environmental effect case and (d) the coalescence rate case. }
  \label{Matrix}
\end{figure*}

  The correlation between the masses of SMBHBs and their host galaxies as well as the redshift evolution of the correlation offers valuable insights into the evolution of SMBH population. The redshift evolution of the $M_{\rm BH}-M_{*}$ relationship can potentially tell us about the accretion and merger history of the SMBHs. If the SMBHs are continuously growing even at low redshifts, we expect to find heavier BHs at low redshifts that reside in similar galaxies than at higher redshifts. This relates to the case with negative $\nu$ where the anisotropy is smaller than the case where $\nu$ is positive.

 \textbf{Environmental effect :} 
 The residence time of a binary at a high frequency, in the absence of environmental effects, is proportional to $f^{-11/3}$. This implies that, compared to sources inspiraling at the lower limit of the PTA band, we can expect the number of sources observable at the higher frequency band of the PTA to be extremely low. 
However, in the presence of environmental effects, the residence time of the binary decreases at lower frequencies. This is because these effects are only significant at large separations of the binary. In a scenario where the galactic environment is more efficient at driving binaries to smaller separations, we would expect to observe a population with a relatively smaller number of binaries at low frequencies than one would expect in the absence of environmental effects.

 In Fig. \ref{env}, we illustrate the mode of $ \Omega_{\rm gw}(f)$ and $C_\ell(f)$ (over 1000 realizations) as a function of frequency for different environmental effects captured by the parameter $\kappa$. Additionally, we depict a 68 \% confidence interval around the mode. In these instances, we maintain a constant total number of binaries. The curves representing $C_{\ell}(f)$ at different $\kappa$ values are very similar to each other with almost the same amplitude B and very little difference in the power law index $\lambda$ of the fitted parameters to the $C_{\ell}(f)$. This is because the environmental effect is important only at low frequencies. At these lower frequencies, the number of sources is much larger, resulting in low anisotropy for both cases. This means that the effect of the spectral change due to the environmental effect is very small on a large-scale distribution of the SGWB.

\textbf{Coalescence rate:} 
At last, in Fig. \ref{z0}, we illustrate the result of $\Omega_{\rm gw}(f)$ and $C_{\ell}(f)$ for different coalescence rates and their evolution. We show the mode of the $\Omega_{\rm gw}(f)$ and $C_{\ell}(f)$ (over 1000 realizations) along with a 68 \% confidence interval around the mode. The parameter $\delta_f$ is the power index of the coalescence rate, and parameters $\mu$ and $z_0$ capture the redshift evolution of the coalescence rate. In the case where $\mu$ is non-zero, we have a redshift-dependent coalescence rate. The parameter $z_0$ represents the redshift below which the coalescence rate of the binaries (see Eq. \ref{Flow_eq}) is higher at lower frequencies. For $z_0 = 1$, the coalescence rate is higher at lower frequencies up to a redshift of 1. As the value of $z_0$ increases, the number of binaries emitting at high frequencies decreases at lower redshifts.

For the case where $z_0$ = 1,  most of the sources are sampled from the coalescence rate distribution that decreases with frequency. In Fig. \ref{dNdf}, we show the number of sources emitting in each frequency bin ($\Delta f$) and each redshift bin ($\Delta z$ = 0.1), along with the theoretical curve at 3 different redshifts for different scenarios of the coalescence rate case. It can be seen that the slope of the $\frac{dN_f}{df} \Delta f$ decreases with redshift for redshift-dependent scenarios (Fig. \ref{N4} and Fig. \ref{N3}) whereas it remains stable in the redshift independent scenarios (Fig. \ref{N5} and Fig. \ref{N2}) of the coalescence rate, as expected. For the redshift-dependent case, the number of sources at low redshift in high-frequency bins becomes so small that the expected number of sources in that frequency bin at that redshift is below 1. However, the Monte Carlo sampling only returns the discrete integer source. Therefore, the analytical curve lies below the simulated results at high frequencies. The effect of this is also seen in the $C_{\ell}(f)$ for two cases: when $\delta_f = 0$ and $\mu = 0$, and when $\delta_f = 1$, $\mu = 1$ and $z_0=1$, where there is a sudden jump at $f = 2.4 \times 10^{-8} $. Just like in the $\nu = 0.5$ curve in Fig. \ref{Cl_nu}, the expectation value of the number of sources in the low redshift bins, emitting at high frequencies bins, is less than one, but the sampled number of sources can be mostly either zero or one. This sporadic nature of the sampling at high-frequency bins results in a sharp increase in the anisotropy at high frequencies.

The fitted curves for $C_{\ell}(f)$ show a significantly larger power law index ($\lambda$) for the redshift-dependent case compared to the redshift-independent case for the same $\delta_f$ value. This difference arises because the number of sources in the redshift-dependent case decreases as the frequency increases. Similarly, the power law index is larger for $\delta_f = 0$ than that of $\delta_f = 1$ for the redshift independent case because there are fewer sources at high frequencies for the former case than for the latter.

The redshift evolution of the coalescence rate can help us understand the merger history of the galaxies. A consistent coalescence rate across redshifts suggests a stable merger rate of galaxies over time. Conversely, if relatively more binaries are found at lower frequencies at some redshifts, it could indicate a higher merger rate during those periods. In the case where the coalescence rate at low frequency (compared to high frequency) increases with redshift, signifies a larger galaxy merger rate at high redshifts.

\textbf{Variation of $C_{\ell}$ with spatial frequencies ($\ell$) and spectral frequencies($f$):}
In Fig. \ref{Matrix} we depict the spectral and spatial structure of the anisotropies in the SGWB for a few different models. The figure shows the mode of $C_{\ell}(f)$ over 1000 realizations as a function of the multipole moment ($\ell$) and frequency ($f$) in color.  The results for different variations of the model parameters for each of these cases at a fixed spatial frequency $\ell= 15$ are discussed previously. Here we mainly focus on the additional gain by measuring the signal in $\ell$ as well. The $\ell = 0$ mode signifies the isotropic component ($\Omega_{\rm gw}(f)$) of the SGWB signal. The presence of non-zero higher-order moments implies additional information in the $\ell - f$ space beyond just the $\ell = 0$ (isotropic) moment. The non-zero higher-order moments indicate the presence of anisotropy in the SGWB as a function of GW frequency. As evident from the color matrix plots in Fig. \ref{Matrix}, the $\ell - f$ structure remains nearly invariant with respect to $\ell$ at all frequencies ($f$). This is expected as at these large angular scales, the properties of the host galaxies and the correlation with the n-Hz GW signal will be similar for a statistically isotropic and homogeneous Universe.

In Fig. \ref{MT1}-- Fig. \ref{MT9}, we display the $\ell-f$ plot of the $C_{\ell}(f)$ for four different cases. Consistent with the behavior at $\ell=15$, the $C_{\ell}(f)$ curves exhibit an increase as the frequency increases from $f= 2\times 10^{-9}$ Hz to $f= 3\times 10^{-8}$ Hz in nearly a steady way at all values of $\ell$. The steady increase at all $\ell$ with frequency $f$ makes the theoretical signal strongly correlated and predictable. As the Universe is expected to be statistically isotropic and homogeneous at large scales, the SGWB fluctuations at large scale are nearly independent of spatial scale $\ell$ values and vary strongly with the spectral frequency $f$ of GWs, which captures the evolution of the SMBHBs. If one can probe angular scales corresponding to galaxies (higher $\ell$s), the spatial behavior is expected to show some variation. As the current PTA analysis can only explore large angular, we have not explored the signal at small angular scales in this paper. The small angular scale signal will be explored in a future work. 

The presence of anisotropic power spectrum with more power at high $f$ for every value of spatial frequency $\ell$ makes it possible to measure the amplitude $B$ and the power-law index $\lambda$ (of a two-parameter model to capture the anisotropic power spectrum) of $C_{\ell}(f)$ by combining different values of $\ell$, which gives a $\sqrt{f_{\rm sky}(2\ell+1)}$ independent information per $\ell$ for a survey with sky-fraction $f_{\rm sky}$. The theoretical analysis presented in this paper demonstrates how the anisotropic signal of the SGWB in the $\ell-f$ space can capture the imprints of SMBHBs evolution in the Universe. We have shown the variation of $C_{\ell}(f)$ for different scenarios of astrophysical models and have theoretical predictions on the scaling of the anisotropic signal with $f$. In future work, we will explore the measurability of this signal and how it can be constrained from the SGWB observations.

A summary of the key signatures of SMBHB evolution on the SGWB spatial-spectral frequency is as follows.
\begin{itemize}
    \item Even in the absence of any redshift evolution of the SMBH mass and galaxy stellar mass relation, the anisotropy in SGWB captured in terms of the spatial-spectral power spectrum $C_{\ell}(f)$ can be large if SMBH masses and galaxy stellar mass relation is more steep implying that heavier SMBHs can be efficiently hosted in galaxies of lower stellar mass. The amplitude and shape of SGWB power spectrum $\Omega_{\rm gw}(f)$ can be comparable for both steep and shallow SMBH masses and galaxy stellar mass relation. 
     \item If heavier SMBHs form efficiently at high redshift, then the amplitude of the anisotropy can increase in the high frequencies of n-Hz signal than the low frequency for a comparable SGWB power spectrum $\Omega_{\rm gw}(f)$. The spectral shape of $\Omega_{\rm gw}(f)$ and spatial-spectral anisotropy power spectra $C_{\ell}(f)$ bring complementary information on the redshift evolution of the SMBH and galaxy stellar mass relation.   
    \item The effect of the environment hardening on the anisotropy of SGWB in the spatial-spectral domain is not very pronounced at large angular scales (larger than one-degree angular scales) as the environment effect is more dominant at small angular scales (less than arc-minutes angular scales). However, the spectral shape of $\Omega_{\rm gw}(f)$ shows prominent features in the presence of environmental effects.
    \item The coalescing rate of SMBHBs across cosmic time shows prominent features in both power spectrum $\Omega_{\rm gw}(f)$ and spatial-spectral anisotropy power spectrum $C_{\ell}(f)$. If the coalescence rate is less efficient in high frequency than in low frequency, and/or is less efficient at low redshift, then the amplitude of $C_{\ell}(f)$ increases. By measuring $\Omega_{\rm gw}(f)$ and $C_{\ell}(f)$ the frequency and redshift evolution of the coalescing rate can be inferred using the n-Hz signal.
\end{itemize}

Measuring the temporal fluctuation of the SGWB and estimate R(z) and mass distribution using temporal fluctuation of the SGWB .

\section{Conclusion}\label{conc}

The study of SMBH formation and evolution remains a central challenge in modern cosmology. The observational study of the SMBH population is constrained by our ability to detect and resolve the source at high redshifts using electromagnetic probes. The detection of n-Hz gravitational waves through radio observations of signals from pulsars presents a promising avenue for unraveling the mysteries surrounding SMBH formation and evolution. The SGWB density from PTA will provide a new approach to understanding the population of SMBHs. 

However, to truly comprehend the SMBH population and its evolutionary trends, it is important to explore additional observational avenues and develop comprehensive theoretical frameworks that can account for the diverse range of processes involved in their formation and growth. Our study has highlighted the importance of studying the isotropic as well as the anisotropic SGWB density in the study of the evolution of SMBH which can shed light on the formation scenarios of SMBHBs. The study of SGWB anisotropy holds great promise as a valuable tool for advancing our understanding of some of the open questions regarding the formation and evolution of SMBHs and SMBHBs. The evolutionary history of the SMBHs is expected to have an imprint on the correlation between the SMBH mass and the properties of the host galaxies. To capture the physics from multi-scale dynamics of the SMBHB population, we adopted a multiscale adaptive technique to cross-match the physics involved at different scales from cosmological to the sub-pc scale where the coalescence of the binary occurs, to calculate the large-scale spatial and spectral anisotropy of the SGWB in n-Hz range. By utilizing large-scale simulations such as \texttt{MICECAT} and smaller-scale, high-resolution simulations such as \texttt{ROMULUS25}, along with an analytical model for the physics of the sub-pc SMBH environment, we demonstrate the impact of SMBHB evolution on the SGWB signal.

Our results have illustrated how the anisotropy of the SGWB from the population of the SMBHBs is affected by the  SMBH mass-stellar mass relation. Similarly, we have also demonstrated that the coalescence rate and its evolution affect the overall shape of the spectral dependence of the angular correlation $C_{\ell}(f)$ of the SGWB. The coalescence rate of the SMBHBs at different separations of binaries can be used as a probe to the merger history of the galaxy. Furthermore, our investigation highlights the significant impact of environmental effects on the anisotropy of the SGWB. The presence of environmental effects, particularly at large separations of the binary, can alter the residence time of a binary at low frequencies. This effect leads to an increase in the anisotropy.

The efficiency of the multiscale adaptive technique can be hindered by the poor resolution and volume of the simulations. Lower resolutions can impact the accuracy of inferred properties, particularly for low-mass sources. Moreover, poor resolution can also affect the modeling of physical processes within these structures, such as star formation, feedback from supernovae and active galactic nuclei, and the growth of SMBH. These processes are crucial for understanding the observed properties of galaxies and SMBHs and their evolution over cosmic time.

Although with the present n-Hz data, the sky localization is poor, the future inclusion of the Square Kilometre Array (SKA)  will significantly improve the sky localization error, as a large number of pulsars can be timed with SKA \citep{lazio2013square, 2018RSPTA.37670293S}. SKA is projected to find several hundred pulsars, which will enhance the PTA's ability to detect and localize the GW signal. With the addition of hundreds of pulsars, SKA is expected to enhance the localization of the GW signal to a few tens of square degrees \citep{sesana2010gravitational,sesana2010measuring}. 
 
The approach proposed in the work will help us put some valuable constraints on the population and evolution of the SMBHs. Using the multiscale adaptive technique employed in this work, we can efficiently calculate the spatial-spectral anisotropy of the SGWB signal for a wide range of astrophysical scenarios of SMBH formation and evolution, which can be tested with n-Hz GWs. In the future, incorporating this framework into a Bayesian data analysis setup will enable the measurement of parameters (such as $\rho, \eta, \nu, \delta_f,\kappa, \mu, z_0$) by using the spatial-spectral anisotropic signal, $C_{\ell}(f)$. Furthermore, cross-correlation of the SGWB signal with the tracers of the large-scale structures (such as galaxies, quasars, AGN) will bring additional insights into the formation of the SMBHs in the Universe. A work is currently under preparation to explore these aspects. Though this study strictly limits the power spectrum of the SGWB, signatures beyond the power spectrum, such as non-Gaussian signatures are also expected to be important due to the sporadic spatial distribution of the SMBHBs in the Universe. These will also bring more constraining power in measuring the SMBHB's evolution parameters.

\section*{Acknowledgments}
This work is a part of the $\langle \texttt{data|theory}\rangle$ \texttt{Universe-Lab} which is supported by the TIFR and the Department of Atomic Energy, Government of India. 
 The authors would like to thank the computer center HPC facility at TIFR for providing computing resources. The authors would also like to acknowledge the use of the following Python packages in this work: Numpy \citep{van2011numpy}, Scipy \citep{jones2001scipy}, Matplotlib \citep{hunter2007matplotlib}, Astropy \citep{robitaille2013astropy,price2018astropy}, Healpy \citep{Zonca2019}, and Ray \citep{moritz2018ray}.

 \section*{Data Availability}
The data underlying this article will be shared at the request to the corresponding author.

\bibliographystyle{mnras}
\bibliography{main_paper}

\setlength{\parskip}{1.8em}

\appendix
\section{Fitting power law to the 15-year data release of NANOGrav}
\label{sec:appendixA}

\begin{figure}
    \centering
    \includegraphics[width=7cm]{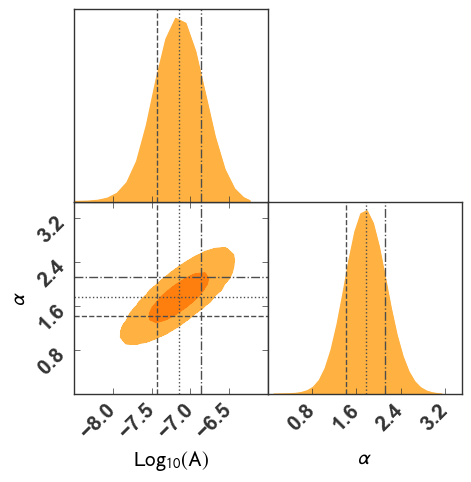}
    \caption{ Corner plot showing the posterior distributions of $\rm Log_{10}(A)$ and $\alpha$ for the power-law model fitted to $\Omega_{\rm gw}(f)$ estimates from NANOGrav 15-yr data set.}
    \label{Corner}
\end{figure}

We fit the $\Omega_{\rm gw}(f)$ of the NANOGrav 15-yr data set to a power law using PTArcade code \citep{mitridate2023ptarcade}.

\begin{equation}
    \Omega_{\rm gw}(f) = A \times (f/f_{yr})^{\alpha},
\end{equation}
where A is the amplitude of the $\Omega_{\rm gw}(f)$ at $f = f_{yr}$, $f_{yr}$ = 1/yr, and $\alpha$ is the power law index. Fig. \ref{Corner} shows the corner plot of A and $\alpha$ fitted to the NANOGrav 15-yr data set. The median values along with the uncertainties on parameters based on the 16th, and 84th percentiles are $\rm Log_{10}$A= $-7.14_{-0.29}^{+0.28}$ and $\alpha$= $1.76_{-0.34}^{+0.36}$. 

In the cases: SMBH mass-Stellar mass and SMBH mass-Stellar mass evolution, we normalize the number of sources contributing to the signal in each frequency bin such that the mode of $\Omega_{\rm gw}(f)$ (over $10^{3}$ realizations) agrees with the power law curve obtained by taking the median values of the parameters A and $\alpha$.

\label{lastpage}
\end{document}